\documentclass[hidelinks,12pt,authoryear]{article}
\usepackage{float}
\floatstyle{plaintop}
\restylefloat{table}
\usepackage{physics}
\usepackage[caption = false]{subfig}
\usepackage[english]{babel}
\usepackage{textcomp}
\usepackage{amssymb}
\usepackage{longtable}
\usepackage{multirow}
\usepackage{mathtools}
\usepackage{amsmath}
\usepackage{amsthm}
\usepackage{amsbsy}
\usepackage{bbm}
\usepackage[font=footnotesize]{caption}
\usepackage{eurosym}
\usepackage[doublespacing]{setspace} 
\usepackage{inputenc}
\usepackage{geometry}
\usepackage[makeroom]{cancel}
\usepackage{setspace}
\usepackage{multirow}
\usepackage{xy}
\usepackage{float}
\usepackage{eurosym}
\usepackage{natbib}
\usepackage{footmisc}
\usepackage{array,ragged2e}
\usepackage{threeparttable}
\usepackage{caption}
\usepackage{bbold}
\usepackage{amsmath}
\usepackage{amsthm}
\usepackage{mathtools}
\usepackage{amsbsy}
\usepackage[pdftex]{color,graphicx}
\usepackage{graphicx}
\usepackage{inputenc}
\usepackage{geometry}
\usepackage{booktabs}
\PassOptionsToPackage{hyphens}{url}
\usepackage{hyperref}
\usepackage{nameref}
\restylefloat{table}

\newcommand{\R}{\mathbb{R}}
\newcommand{\e}[1]{{\mathbb E}\left[ #1 \right]}

\newcommand\independent{\protect\mathpalette{\protect\independenT}{\perp}}
\newcommand{\ou}{
  \mathrel{
    \vcenter{\offinterlineskip
      \ialign{##\cr$<$\cr\noalign{\kern-1.5pt}$>$\cr}%
    }
  }
}

\makeatletter
\newcommand\footnoteref[1]{\protected@xdef\@thefnmark{\ref{#1}}\@footnotemark}
\makeatother

\def\independenT#1#2{\mathrel{\rlap{$#1#2$}\mkern2mu{#1#2}}}
\newcommand{\argmin}{\operatornamewithlimits{argmin}}
\newgeometry{
 left=3.175cm, 
  right=3.175cm,
  top=3.175cm,
  bottom=3.175cm,
  bindingoffset=5mm
}

\begin{document}

%%%%%%%%%%%%%%%%%%%%% TITLE
\begin{titlepage}

\title{Policy evaluation of waste pricing programs using heterogeneous causal effect estimation}

\author{Marica Valente\thanks{\scriptsize University of Innsbruck (marica.valente@uibk.ac.at) and DIW Berlin.\ This paper, partly written at DIW Berlin and ETH Zurich, was previously circulated under the title ``Heterogeneous effects of waste pricing policies". I thank my advisors Bernd Fitzenberger and Jeffrey Wooldridge for their invaluable advice. I also thank Pio Baake, Lucas Bretschger, Marco Caliendo, Juan Carlos Escanciano, Elia Lapenta, Julien Sauvagnat, Stefan Seifert, Jann Spiess, and Lorenzo Trapani for fruitful discussion. I am grateful for seminar invitations and feedback received from participants of the NBER Environmental \& Energy Economics Workshop, LSE Causal Inference Meeting, University Carlos III of Madrid, CREST, Aix-Marseille University, University of Innsbruck, ECB, Bank of Spain, Trinity College Dublin, Mines ParisTech, University of the Balearic Islands, University of Nottingham, University of Barcelona, FBK and University of Trento, LMU Munich, University of Hohenheim, University of Ferrara, Berlin Research Institute for Global Commons and Climate Change, University of St.Gallen, Institute for Employment Research of Nuremberg, DIW Berlin, Berlin Network of Labor Market Research, University of Southern Denmark, University of Potsdam, the Lindau Nobel Laureate Meetings, and the ETH/UZH Seminars in Economics \& Data Science. This work did also benefit from presentations in the (2022) Women in Empirical Microeconomics Conference in Chicago, (2021) IAERE Annual Conference (Best Paper Award), ESEM Annual Congress, New York Camp Econometrics, (2020) ES World Congress, EAERE Annual Conference, NERI Workshop in Naples, Causal Machine Learning Workshop in St. Gallen, and (2019) ES Winter Meeting, European Conference of the Society for Benefit-Cost Analysis in Toulouse, EEA Annual Congress, EAERE Annual Conference, IAERE Annual Conference, and EUI Annual Conference. The author declares that she has no relevant or financial interests that relate to the research in this paper.} }

\date{} 

\maketitle 	

\vspace{-0.8cm}

\noindent \textbf{Abstract}\ 
Using machine learning methods in a quasi-experimental setting, I study the heterogeneous effects of introducing waste prices---unit prices on household unsorted waste disposal---on waste demands and municipal costs. Using a unique panel of Italian municipalities with large variation in prices and observables, I show that waste demands are nonlinear. I find evidence of constant elasticities at low prices, and increasing elasticities at high prices driven by income effects and waste habits before policy. The policy reduces waste management costs in all municipalities after three years of adoption, when prices cause significant reductions in total waste.

\vspace{0.5cm}

%https://www.aeaweb.org/econlit/jelCodes.php?view=econlit&print
\end{titlepage}

%%%%%%%%%%%%%%%%%%%%% INTRODUCTION

\newpage

\section*{Introduction} \label{Sec:intro}

Waste management generates pollution externalities that are not internalized by households under traditional flat pricing schemes \citep{Fullerton1996}. To correct this inefficiency, a growing number of municipalities have implemented Pay-As-You-Throw (PAYT) programs that require households to pay for each unit (per bag, can, or weight) of unsorted waste presented for collection.\footnote{PAYT is used in more than 800 U.S. cities, and is broadly implemented in Europe, Japan and Korea. China planned a number of pilots in big cities as Beijing \citep{CM2015}.} Policy average causal effects on waste quantities can be estimated, e.g., with fixed effects regression methods and synthetic controls.\footnote{See, e.g., \citet{Valente2019} and the survey in \citet{Bel2014}.} However, empirical estimates provide mixed evidence on the magnitude of PAYT average effect on unsorted waste as well as on possible indirect causal effects on recycling and total waste. Moreover, policy impacts may vary across municipalities depending, e.g., on the adopted price level and socio-economic characteristics \citep{Kinnaman2006}. In light of this, heterogeneity in causal effects plays an essential role in evaluating policy efficacy, allowing to ascertain subpopulations for which the policy is most beneficial and to generalize estimates to a new target population. On top of this, heterogeneous causal effect estimates can be combined with municipal waste management costs to deliver heterogeneous cost-benefit evaluations of waste policies.

In this paper, I examine heterogeneous demand responses to waste prices, and their impact on municipal waste management costs. The main challenge in the analysis is that determinants of waste generation and policy adoption are possibly many, and may confound the estimation of causal effects.\footnote{\label{foot2} Previous studies suggest that waste generation determinants such as income, education and pre-policy waste levels may also drive policy adoption decisions \citep[see e.g.][]{Gradus2019}.} This motivates the collection of a unique panel of municipalities with a large variation in prices and observables, and the estimation of municipal level causal effects of prices (continuous treatment) via machine learning methods. These techniques, in fact, allow to control for a high-dimensional set of covariates.

I estimate policy effects on municipal costs for each municipality by combining price causal effects on unsorted and recycling waste with their impacts on waste management costs. I find that waste prices decrease municipal costs for all municipalities especially in the long-term (after three adoption years) when municipalities show the highest reductions of total waste. 

This paper aims to contribute to the causal inference literature. I use machine learning methods to estimate heterogeneous causal effects when units are heterogeneous not only in their observable characteristics but also in their treatment level. I provide a simple approach to interpret causal effect heterogeneity in the treatment level, and to exploit unit level estimates for policy targeting and cost analysis. Also, I apply machine learning methods in such a way to account for a high-dimensional set of possible sources of self-selection and for staggered policy adoption. In addition, while several studies applying machine learning to binary treatment effect models can be found in labor economics,\footnote{See, e.g., \citealt{athey2019,Davis2017}.} applications of high-dimensional methods for causal inference in environmental economics are scarce \citep{christensen2021, Priest2020}.

My work also adds to a large and growing literature on waste prices. Prior research estimates the average price elasticity of waste demands or the average causal effect of PAYT (binary treatment) on waste amounts.\footnote{See, e.g., \citealt[]{Valente2019, carattini2018, Bucciol2015, Huang2011, Allers2010, Fullerton2000, Fullerton1996}. See \citealt{Kinnaman14} for a review.} Compared to these studies, this paper requires less restrictive identifying assumptions, in particular, allowing for possibly heterogeneous price effects across units (heterogeneity in prices and observables). Further, it aims to deliver more precise estimates as it exploits a large variation in prices and flexibly accounts for many possible sources of self-selection. 

Beyond the methodological contribution, this paper looks for the first time at demand responses at high prices, providing new evidence on PAYT effectiveness.\footnote{My data include prices between \EUR{}0.01/L and \EUR{}0.18/L. High prices are defined by values above the median of \EUR{}0.09/L. The literature has so far looked at demand responses at low prices. The highest price in, e.g., \cite{Fullerton2000} is \$2.18 per 32-gallon bag (120L) or \$0.02/L. See \citet{Bel2014} for a survey.} Previous studies typically estimate low price elasticities of waste demands \citep[for a review see, e.g.,][]{Kinnaman2006}. In line with prior work I find low elasticities for municipalities adopting low prices. However, for municipalities adopting high prices, I find significantly larger elasticities. Moreover, at high prices waste demands also respond to other factors, in particular, elasticities are largest for municipalities with low income and little recycling before policy. Finally, this is the first study to perform a waste management cost analysis at municipal level, showing significant cost savings from PAYT adoption for most municipalities.

I study household waste generation behaviors of about 3,600 Italian municipalities over 2010-2015. Italy provides an ideal setting to study heterogeneous effects of waste prices because both price levels and socio-economic characteristics largely vary across municipalities. I use web scraped and administrative data to construct a new and rich dataset on waste generation and price adoption at the municipal level. The final dataset includes 45 different price levels ranging from 1 to 18 euro (\EUR{}) cents per liter of unsorted waste, and 90 municipal characteristics that may explain price adoption and waste generation. 

To consistently estimate municipal level parameters with high-dimensional data, I use machine learning-inspired matching estimators called generalized Random Forests (RFs) \citep{atheyGRF}.\footnote{RFs build upon \cite{atheyRF}, \cite{atheyimbens2016} and, firstly, \cite{Breiman2001}.} Intuitively, RFs partition the large covariate space into small neighborhoods of municipalities mostly similar in those characteristics that drive parameter heterogeneity.  Within neighborhoods, I estimate constant treatment effects by the residual-on-residual regression estimator, or R-learner \citep{nie2019}.\footnote{Robust to confounding affecting outcome and treatment \citep{Chernozhukovetal2017}.} The random forest instantiation of the R-learner allows to flexibly control for a high-dimensional set of outcome determinants and sources of self-selection. Further, thanks to data subsampling and aggregation, random forests can be adapted to account for staggered adoption designs.\footnote{See implementation details in Section \ref{Sec:results}.}

The advantage of using RFs is to relax the assumption of constant price effects across municipalities and estimate the full distribution of price effects for all municipalities with pointwise-consistent confidence intervals. Estimation avoids ad hoc modeling choices, and flexibly allows for parameter heterogeneity in the large set of (often correlated) covariates.\footnote{Differently, standard regression methods are justified if treatment effects are constant, observables have linear or pre-specified effects, and unobservables are time-invariant.} The improvement of RFs vis-\`a-vis, for instance, theory-informed heterogeneity analysis is to provide a data-driven documentation of heterogeneous causal effects, as opposed to specification search.

To guide my empirical investigation, I formulate a simple model of household waste disposal choices
building on the seminal work of \cite{Kinnaman17book}. The representative household generates waste from consumption, and chooses to spend some time and effort in recycling. The household maximizes utility from consumption subject to the budget constraint, by choosing the quantity of material to discard and to recycle. Comparative statics characterize how, after the introduction of PAYT, the amount of unsorted, recycling, and total waste varies with household characteristics and waste generation habits. The model delivers two key theoretical predictions. First, after PAYT introduction, households are predicted to reduce unsorted waste, through both an increase in recycling and a reduction in total waste. Second, PAYT effects are heterogeneous across municipalities depending on, in particular, income and waste generation habits. Therefore, due to the heterogeneity in PAYT effects on municipal household waste, the sign and magnitude of PAYT effects on municipal waste management costs is ambiguous because it depends on whether cost savings from unsorted waste reductions exceed cost increases from higher recycling. The size and richness of my data, coupled with the use of machine learning methods, allows to characterize the heterogeneity in responses at the municipal level, and provide novel insights into the mechanism driving the response of waste behavior to PAYT.

I start by estimating municipal level price elasticities of demands for unsorted, recycling, and total waste per capita. The hypothesis of no heterogeneity is rejected for all outcomes. %Waste prices cause large unsorted waste reductions driven by increased recycling and, to a smaller extent, less total waste. 
To disentangle sources of heterogeneity and interpret municipal level causal estimates, I regress these elasticities on price levels and a parsimonious set of relevant regressors capturing household costs of waste disposal: income, education, and pre-policy waste levels.\footnote{Their relevance is discussed in, e.g., \cite{Valente2019} and \cite{GilliBook2018}.} I find a nonlinear relationship between elasticities and price levels: while at high prices (above $9$ cents) elasticities are increasing, elasticities are rather constant at low prices. At high prices, a one cent price increase reduces unsorted waste by 5 to 10\% (vs. 4.7-5.7\% at low prices), increases recycling by 2 to 6\% (2.5-3.2\%) and reduces total waste by 0.1 to 0.7\% (0.6-0.8\%) compared to a counterfactual zero-price (no PAYT) scenario. Analyzing price indirect effects on recycling and total waste, I find that higher prices do not cause further total waste reductions, instead they simply reallocate waste to the recycling pile.

%Having established heterogeneity of price effects, the second part of the empirical analysis focuses on evaluating policy effects on municipal waste management costs. I estimate that PAYT leaves unit costs of waste mostly unaffected, suggesting constant returns to scale.\footnote{This is consistent with results on Italian municipalities by, e.g., \cite{Abrate2014}.} In a small share of municipalities, I estimate that PAYT causes an increase (decrease) of unit costs of unsorted waste management (recycling). Municipal private benefits of recycling from selling the fraction of recycled material are included in the analysis. 

Next, I simulate the impact of PAYT adoption on municipal waste management costs combining predicted (causal) changes in unsorted and recycling waste for each municipality with unit costs of waste management under PAYT. Municipal private benefits of recycling from selling the fraction of recycled material are included in the analysis. After three years of adoption, I predict annual cost savings for all municipalities of on average \EUR{}24 per capita, however, there is large variation. As unit costs of unsorted waste are higher than those of recycling for most municipalities, total waste reductions trigger large reductions in waste management costs. This implies that also low prices -- by causing significant total waste reductions -- improve waste behavior and municipal budgets substantially. 

I present additional analyses that show robustness of my results against the presence of unaccounted-for heterogeneity. As often found in the literature \citep[see][for a review]{Bel2014}, my results are robust to confounding from adoption of heterogeneous PAYT systems (weight versus volume). Further, I find that waste tourism or, more generally, spillover effects are not relevant on any significant scale.\footnote{Waste tourism is a way to evade PAYT systems by discarding waste in locations where unit fees are not applied.} Finally, I contrast my average estimates to the binary treatment case, two-way (event-study-like) fixed effects estimation (with/without re-weighting á la \citealt{Chaise2020}), and R-learning LASSO regression. Results highlight the importance to account for continuous rather than binary treatment, the bias of the fixed effects estimator due to non-parallel trends pre-policy, and robustness to the specific choice of R-learning estimator.

The remainder of the paper is structured as follows. Section \ref{Sec:data} describes policy background and data. Section \ref{Sec:theorymethod} discusses the empirical framework.  Section \ref{Sec:results} presents the main results and their policy implications. Section \ref{Sec:conclusions} concludes.

	\section{Background and Data} \label{Sec:data}

\subsection{PAYT policies} \label{Sec:policy}
PAYT policies in Italy, as in many municipalities worldwide, require households to pay a price per unit of unsorted waste according to either its volume (per bag or bin) or weight (per kilogram). PAYT fulfills the equivalence principle in the sense that waste service consumers pay for its consumption (as, e.g., for energy and water), and the polluter-pays principle for which households pay according to their unsorted waste. 

The baseline policy in both PAYT and non-PAYT municipalities is a flat fee independent of waste quantities, namely, the unit price is zero. Flat fees depend on house (m$^2$) and household size (number of inhabitants). PAYT municipalities use the flat fee to cover only fixed costs of waste management, and implement waste prices to cover variable costs. 

Municipalities can decide whether and when to implement PAYT, as well as the price level and collection system. Policy adoption decisions are based on, for instance, goals of waste pollution and management cost reduction. Yet, political and socio-economic factors may also matter (see \citealt{Gradus2019} and Section \ref{Sec:sel} for details). In sum, drivers of price adoption and waste generation often overlap, and are possibly many. Their relevance is, therefore, an empirical question.

Credible enforcement and monitoring systems are crucial for policy success.\footnote{E.g., trash bins are locked, drones and photo traps track illegal dumping, and reciprocal monitoring is enforced by charging all households in a building for lack of policy compliance \citep{CONSEA2019}.} Illegal dumping is the main possible adverse effect. However, after about fifty years of PAYT experiences worldwide, adverse effects seem a bigger fear than reality.\footnote{See, e.g., \cite{Valente2019} for Italy and \cite{Skumatz2008} for the US. The latter found illegal dumping to be a short-term issue which lasts three months or less and involves 3\% of municipalities.} In Italy, anecdotal evidence suggests that (i) waste tourism in surrounding municipalities is a rare and short-lived phenomenon, (ii) enforcement and monitoring systems allow to actually decrease illegal dumping episodes, and (iii)
PAYT helps meeting goals of increasing recycling, preventing waste, and minimizing waste sent to landfills \citep{Legamb17}.

	\subsection{Data} \label{Sec:dat0}
	
Using web scraped and administrative data, I construct a new municipal level dataset with annual information on waste prices, amounts of household waste (main outcomes) and management costs, as well as socio-economic, geographic, and political determinants of waste generation and price adoption. The final database is a panel of Northern and Central Italian municipalities over the sample period 2010-2015. I exclude municipalities with missing values as well as South and Insular Italy because arguably not suitable for being in the control group.\footnote{Missing values are due to merging administrations and data errors. In South and Insular Italy, as defined by the NUTS 1 classification, there are no treated municipalities over the sample period, and data is incline to significant mismeasurement due to illegal disposal \citep{ISTATcrime}.}  Moreover, outcome data of treated municipalities must be observable at least two years before policy adoption.\footnote{PAYT is typically announced one year before implementation. In this year, policy anticipation effects may influence waste outcomes. Thus, outcomes of at least two years before policy are needed to predict waste behavior and adoption decisions without bias.}

\begin{figure}[H]
        \centering
\subfloat {\includegraphics[trim={0cm 1cm 0 0},clip, scale=0.5]{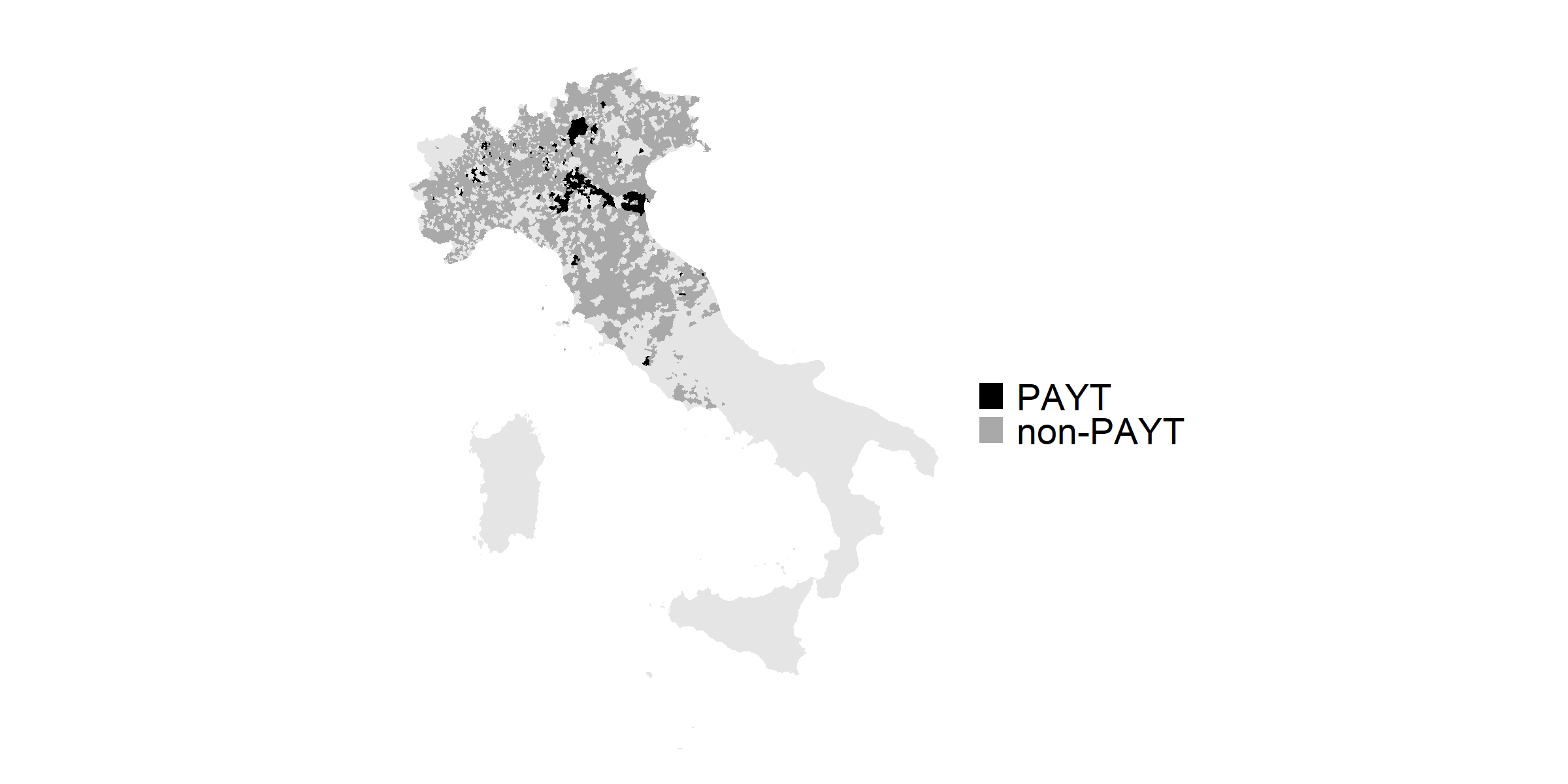}}
      \caption{ \footnotesize Map of PAYT and non-PAYT municipalities in the sample. White areas show the excluded South and Islands, municipalities with missing values, and those adopting PAYT before 2012.} 
      \label{fig:map_dynamic}
    \end{figure}

 The resulting municipalities are 3,574. Waste prices, the treatment, cover 1.7 million people living in 194 municipalities, of which 106 are located in the North-West, 82 in the North-East, and six in the Center. Most of the treated municipalities implement PAYT for the first time in 2013 (77), while the others followed in 2012 (48), 2014 (36), and 2015 (33). Figure \ref{fig:map_dynamic} shows the distribution of PAYT and non-PAYT municipalities in the sample.

\vspace{-0.5cm}
\noindent
This data allow for (counterfactual) predictions of policy effects for 26 million people (45\% of the population) living in 3,380 municipalities without PAYT.

Variables come from a variety of sources. Waste generation and management costs are from the National Environmental Protection Agency (ISPRA). Municipal socio-economic attributes are from  the  National  Institute  of  Statistics  (ISTAT)  and  web scraping  the  online database of comuni-italiani.it.  Political variables on, e.g., municipal elections and mayors' characteristics are obtained from the Ministry of the Interior. Geographic variables measuring, e.g., the distance to waste treatment sites are geocoded using the software R and data of the European Pollution Release and Transfer Register (E-PRTR).

At the time of writing, there is no open-source database on PAYT prices in Italy. I acquired the list of PAYT municipalities from the National Environmental Protection Agency (ISPRA), the National Association of Italian Municipalities (ANCI), and from waste management companies.\footnote{E.g., Aimag, Ascit, Clara, Dolomiti Energia, Hera, Iren,  SEAB, Soraris, etc.} I collected a cross-section of price data directly from municipalities and waste companies, and indirectly from municipal balance sheets. Each municipality sets the same price over the three post-policy years analyzed in this study. Variables are described in Appendix \ref{Sec:appendixA}.

	\subsection{Summary statistics} \label{Sec:sumstat}

Table \ref{tab:nevertreated} compares attributes of never-treated (non-PAYT) and treated (PAYT) units before policy.\footnote{To simplify the exposition, variable values are averaged over all years but interpretation of summary statistics does not change when values are averaged over individual years.} The main outcomes are yearly kilograms of per capita unsorted (UW), recycling waste (RW), and total (TW) which is the sum of the previous two. Municipal socio-economic characteristics such as average household size, income per capita, social deprivation index, commuting intensity, and migration flows are based on annual population data. Table \ref{table:var_desc} in Appendix presents a detailed description of all variables and units of measurement. The relevant groups of predictors are discussed in the next section.

\newpage

\renewcommand{\arraystretch}{1}
\footnotesize
\begin{longtable}{llll|ll} \centering
 \endfirsthead
 \endhead
    \hline \multicolumn{6}{r}{{Cont'd}} 
\endfoot
\endlastfoot
\caption{\footnotesize Summary statistics for never-treated vs. treated municipalities before policy. Observations equal 18,812 for never-treated (unbalanced panel) and 636 for treated before policy (balanced). Data includes 44 lag values of outcomes and costs for a total of 90 covariates.}
\label{tab:nevertreated} \\
\footnotesize
  &   &   \multicolumn{2}{c|}{\underline{Never-treated}}&  \multicolumn{2}{c}{\underline{Treated (before)}} \\
{Obs.} 19,448  &   &   Mean & Sd &  Mean & Sd \\

{Outcomes:} & Recycling Waste (RW) kg p.c. & 233.9 & 82.84 & 296.8 & 89.9  \\ 
 & Unsorted Waste (UW) kg p.c. & 223.5 & 120.5 & 216.0 & 115.1 \\ 
 &  Total Waste (TW=RW+UW) & 457.4 & 127.0 & 512.7 & 142.9  \\ 
   \hline
 
{Costs \EUR{}:} &   RW unit costs per kg & 0.18 & 0.12 & 0.16 & 0.08 \\ 
&  UW unit costs per kg & 0.28 & 0.19 & 0.29 & 0.16 \\ 
&  RW management costs p.c. & 40.34 & 22.48 & 47.73 & 24.31  \\ 
&  UW management costs p.c. & 53.83 & 34.26 & 50.01 & 23.88  \\ 
  \hline

 Socio-economic & Population density & 337.29 & 579.97 & 286.47 & 406.30 \\ 
covariates:&  Household size & 2.31 & 0.34 & 2.50 & 0.72 \\ 
&  Income p.c. (x \EUR{}1,000) & 13.86 & 2.29 & 14.22 & 1.97 \\ 
&  Net migration p.c. & 0.003 & 0.01 & 0.02 & 0.64 \\ 
&  Population (x 1,000) & 7.35 & 34.72 & 9.74 & 21.20 \\ 
&  Share of foreign population & 0.08 & 0.04 & 0.10 & 0.04 \\ 
&  Share of male population & 0.49 & 0.02 & 0.49 & 0.01 \\ 
&  Population growth & 0.003 & 0.15 & 0.03 & 0.12 \\ 
&  Tourism p.c. & 0.33 & 0.85 & 0.27 & 0.82 \\ 
&  Share of pop. aged $<5$ & 0.05 & 0.01 & 0.05 & 0.01 \\ 
&  Share of pop. aged $<14$ & 0.13 & 0.03 & 0.14 & 0.02 \\ 
&  Share of pop. aged $>65$ & 0.23 & 0.05 & 0.21 & 0.04 \\ 
&  Pop. share w. elementary deg. & 0.31 & 0.05 & 0.31 & 0.04 \\ 
&  Pop. share w. college deg. & 0.09 & 0.03 & 0.09 & 0.03 \\ 
&  Share of rented houses & 0.09 & 0.03 & 0.09 & 0.02 \\ 
&  Housing density & 2.30 & 0.27 & 2.23 & 0.25 \\ 
&  Share of single-parent families & 0.10 & 0.02 & 0.10 & 0.01 \\ 
&  Share of students aged $>15$ & 0.06 & 0.01 & 0.06 & 0.01 \\ 
&  Share of commuters & 0.26 & 0.06 & 0.25 & 0.05 \\ 
&  Social deprivation index & -1.50 & 1.58 & -1.79 & 1.25 \\ 
&   Out of labor force/active pop. & 0.63 & 0.06 & 0.62 & 0.05 \\ 
&  Unemployment rate & 0.07 & 0.02 & 0.06 & 0.02 \\ 
&  Commuting intensity & 0.44 & 0.13 & 4.52 & 12.80 \\ \\

{Geographic} &     Km to PAYT city & 50.00 & 63.23 & 33.75 & 37.97 \\ 
covariates: &  Km to hazardous waste site & 13.80 & 15.40 & 9.37 & 9.15 \\ 
 & Km to incinerator & 29.00 & 28.39 & 21.71 & 19.35 \\ 
 & Km to landfill & 11.21 & 10.95 & 7.50 & 6.34 \\ 
 & Province or region capital (0/1) & 0.01 & 0.12 & 0.04 & 0.19 \\
 & Province capital (0/1) & 0.01 & 0.10 & 0.03 & 0.17 \\ 
  & Region capital (0/1) & 0.003 & 0.06 & 0.005 & 0.07 \\ 
  & Medium urbanization (0/1) & 0.38 & 0.49 & 0.51 & 0.50 \\ 
 & High urbanization (0/1) & 0.20 & 0.40 & 0.16 & 0.37 \\ 
 & Low urbanization (0/1) & 0.42 & 0.49 & 0.33 & 0.47 \\ \\

 {Political} &  Political participation  & 0.69 & 0.09 & 0.64 & 0.21 \\ 
covariates: &  Vote share big-tent party & 0.23 & 0.07 & 0.23 & 0.05 \\ 
&  Vote share extreme left & 0.07 & 0.10 & 0.06 & 0.02 \\ 
&  Vote share extreme right & 0.13 & 0.08 & 0.12 & 0.06 \\ 
&  Local mayor (0/1) & 0.76 & 0.42 & 0.86 & 0.35 \\ 
&  Centre-party mayor (0/1)  & 0.003 & 0.05 & 0.00 & 0.00 \\ 
&  Green-party mayor (0/1)  & 0.01 & 0.11 & 0.02 & 0.15 \\ 
&  Left-party mayor (0/1) & 0.07 & 0.25 & 0.09 & 0.28 \\ 
&  Other-party mayor (0/1) & 0.11 & 0.32 & 0.11 & 0.31 \\ 
&  Local-party mayor (0/1) & 0.71 & 0.46 & 0.64 & 0.48 \\ 
&  Right-party mayor (0/1) & 0.10 & 0.30 & 0.14 & 0.35 \\ 
&  Mayor's age & 52.04 & 10.47 & 50.80 & 10.18 \\ 
&  Mayor's years of office & 1.84 & 1.38 & 1.83 & 1.28 \\
  \hline

\end{longtable}
\renewcommand{\arraystretch}{1.2}
\noindent \normalsize
Municipalities generate relatively more RW than UW on average. Managing one unit of UW costs on average twice as much as RW. However, municipal unit costs are largely heterogeneous, and for some of these municipalities (40\%) unit costs of RW are higher than those of UW. Table \ref{tab:nevertreated} also shows that average costs for UW are slightly higher in per capita terms. Per capita costs indicate that one individual on average spends about \EUR{}100 per year for waste services, and pays the most for unsorted waste.\footnote{Municipal p.c. costs approximate household expenditures as, in Italy, municipalities are cost minimizers such that fee revenues finance total waste management costs, a principle known as budget balance.} This implies that UW reductions that do not translate into RW increases will potentially drive cost savings for both households and municipalities. Previous studies point to education and income levels as key determinants of policy adoption and effectiveness \citep[see, e.g.,][]{Gradus2019, Kinnaman2006}. In my sample, PAYT is implemented in municipalities with on average slightly higher income and comparable education levels.

In order to obtain credible estimates of PAYT causal effects for each price level, all municipalities must have close comparisons in the opposite treatment group (PAYT/non-PAYT) for unbiased estimates of the causal effects without extrapolation \citep{ImbensRubin2015}. Overlap in the estimated (continuous) treatment propensities guarantees that a comparable unit can be found for each municipality.\footnote{When overlap is limited, one may consider trimming the sample and focusing on a subgroup with bounded propensity scores \citep{imbenslimitedoverlap}.} For a continuous treatment, the generalized propensity score (GPS) is the conditional probability of receiving a particular level of the treatment given the covariates \citep{Imbens2000}. Figure \ref{fig:overlap} in Appendix shows that PAYT units are fully in the support of non-PAYT units. Overlap statistics show that treated and control groups are similar in their GPS means, first and third quartiles. Moreover, the (normalized) difference in the estimated GPS amounts to less than one standard deviation (0.62), showing that the two groups are not apart. All municipalities satisfy the overlap assumption and can be included in the analysis. %Overlap statistics show that treated and control groups are similar in their GPS means, first and third quartiles. Two additional statistics are particularly useful to analyze GPS overlap (for more details, see \citealp{ImbensRubin2015}, and \citealp{Kluve2012}). First, I compute the (normalized) difference in the estimated GPS. This difference amounts to less than one standard deviation (0.62), showing that the two groups are not apart. Second, I check whether all units have close comparisons in the opposite treatment group. I find that for all treated units and for $96\%$ of the control units there are units with the other treatment status that have differences in GPS less than 10\%, a threshold that guarantees unbiased estimates of the causal effects without extrapolation \citep{ImbensRubin2015}. Thus, all municipalities satisfy the overlap assumption and can be included in the analysis. %Therefore, causal effects for the control group, and not only for the subpopulation of treated units, can be credibly estimated under unconfoundedness.

%Overlap statistics further show that the two groups are not apart. 

Focusing on treated municipalities, Table \ref{tab:befaf} compares per capita waste amounts and municipal waste management costs before and after policy.

\renewcommand{\arraystretch}{1.2}

\begin{table}[H]
\centering
\caption{\footnotesize Before-and-after comparison of waste outcomes and management costs of treated units. Observations equal 1,164 for 194 treated units observed over all years (balanced panel).}
\footnotesize
\begin{tabular}{llll|ll}
 \multicolumn{1}{l}{} &  \multicolumn{1}{l}{} &  \multicolumn{2}{c|}{\underline{Before}} & \multicolumn{2}{c}{\underline{After}}
 \\ 
 {Obs.} 1,164  &  &  Mean &  Sd &  Mean &  Sd\\
{Outcomes:}  
 & Recycling Waste (RW) kg p.c.  & 296.8 & 89.90 & 337.9 & 92.10 \\ 
&  Unsorted Waste (UW) kg p.c. & 216.0 & 115.1 & 115.9 & 80.90 \\ 
 &  Total Waste (TW=RW+UW)  & 512.7 & 142.9 & 453.8 & 127.2  \\ 
\hline
  
{Costs \EUR{}:} & RW management costs p.c. & 47.73 & 24.31 & 54.29 & 20.20 \\ 
 & UW management costs p.c. & 50.01 & 23.88 & 40.31 & 18.11 \\

 &  RW unit costs per kg & 0.16 & 0.08 & 0.17 & 0.07 \\

 & UW unit costs per kg & 0.29 & 0.16 & 0.43 & 0.26 \\ 
 \hline

\end{tabular}
\label{tab:befaf}
\end{table}
\normalsize
\noindent As expected, UW and TW decrease, and RW increases after PAYT. This leads to higher costs of RW and lower costs of UW per capita. Unit costs of UW increase after policy suggesting possible economies of scale. Yet, noticeably larger standard deviations relative to the mean indicate that average effects mask important heterogeneities across municipalities. 

Table \ref{tab:price123} reports price summary statistics by policy year, namely, for municipalities observed after one, two, and three years after policy adoption. Municipalities adopted PAYT starting from 2012 onward, so policy years range between 2012 and 2015. In addition, I have collected data for 2010 and 2011 to make sure to have at least two pre-policy years for all PAYT municipalities included in the sample. Prices range between 1 cent and 18 euro cents per liter of UW, for a total of 45 price levels. The average price is 8 cents, and the standard deviation is 0.05. Performing Wilcoxon and Kolmogorov-Smirnov tests (\citeyear{Wil1945, Kolmogorov1971}) shows that price distributions do not statistically differ across policy years.

\begin{table}[H]
\centering
\renewcommand{\arraystretch}{1.1}
\footnotesize
\begin{tabular}{llllllll}
{Treatment} & {Year} & Mean & Sd & Min &  P50& P75 & Max \\ 
  \hline
 PAYT price \EUR{} per liter & 1 (obs. 194)  &  0.073 & 0.046 & 0.01 & 0.066 & 0.13 &0.18\\
& 2 (obs. 161) & 0.080 & 0.046 & 0.01 & 0.079 & 0.13 & 0.18\\
& 3 (obs. 125) & 0.088 & 0.044 & 0.01 & 0.090 &0.13 & 0.18\\
\hline
\end{tabular}
\caption{ Price summary statistics for PAYT municipalities by policy year.}
\label{tab:price123}
\end{table}

Using prices and UW post-policy, I calculate household variable costs in each municipality.\footnote{I use waste companies' data for liter-to-kg conversion. If missing, I use the median value.} In the third policy year, households in high-price municipalities ($>$13 cents) pay on average \EUR{}176 per capita. In low-price municipalities ($<$3 cents), they pay on average \EUR{}21 per capita. The magnitude of this gap raises several questions. After partialling out confounding factors, do reactions in municipalities setting high versus low prices differ? What role do income effects play in driving these differences? More broadly, how do behaviors toward recycling and total waste adjust to a decrease in unsorted waste? 

\subsection{Determinants of waste generation and policy adoption} \label{Sec:sel}
\normalsize
A large number of variables (confounders) can explain the decision to adopt PAYT as well as differences in waste generation across municipalities. I control for a large set of observables (deconfounders) that potentially affect household waste generation and policy adoption decisions at municipal level. Municipalities decide to implement PAYT because of possibly many individual or combined factors \citep{Gradus2019}: high levels of waste and/or waste management costs pre-policy, awareness of the waste pollution problem, citizens' demand for higher-quality services, political variables, social norms, information effects, a developed recycling infrastructure and large administrative capacity. Further, as shown in \citet{Valente2019}, leftover unobserved time-varying heterogeneity can be controlled for by including outcome lags. As a result, I include waste generation and policy adoption determinants for a total of 90 municipal attributes which can be grouped into five categories: socio-economic and geographic characteristics, political variables, pre-policy waste amounts and management costs.

\vspace{0.5cm}
\textit{Socio-economic characteristics.}--This group of covariates proxies for heterogeneous household opportunity costs of time and space. Above all, the literature highlights income and education levels.\footnote{See, e.g., \citet{Callan2006, Kinnaman2006, Jenkins2003, Miranda1996, VanHoutven1999, Richardson1978,Grossman1974, Wertz1976}.} The latter is measured by the population share with graduate degree or higher, and elementary degree or none. Income effects are indeterminate a priori \citep{Callan2006}. Higher-income households may be less elastic due to, e.g., lower budget constraints and higher time opportunity costs. Both education and income may proxy for citizens' demands for environmental quality \citep{Dunlap2000}. These variables, therefore, may drive both policy adoption and effects. Further, I control for municipal level demographics as, e.g., average house size and home ownership (based on the 2011 Italian national census), average household size and age (based on annual population data), and tourism intensity (in per capita values based on annual population data). Labor market attributes proxy for, e.g., time spent at home. I include the share of unemployed and out-of-the-labor force population, and indeces of labor market activity, commuting intensity, and social deprivation (based on annual population flows). Table \ref{table:var_desc} in Appendix presents a detailed description of all variables and units of measurement. 

\vspace{0.5cm}
\textit{Geographic characteristics.}--I control for the distance of each municipality to waste incinerators, landfills, and hazardous waste treatment facilities. Proximity to waste sites may induce, for instance, lower waste generation through households' pollution awareness. In addition, distant waste sites may cause high transportation costs and, therefore, correlate with price adoption decisions. I further distinguished communities by  urbanization levels, and regional or provincial seats. These variables proxy for differences in administrative capacity, recycling infrastructure, and PAYT system \citep{Gradus2019}. 

Moreover, municipalities may adopt PAYT to mimic successful neighbors' policies \citep{Allers2010}. In other words, vicinity to other PAYT municipalities may matter because of information dissemination. As a proxy, I control for the distance to the closest municipality that implemented PAYT in prior years.

\vspace{0.5cm}
\textit{Political variables.}-- Policymakers may adopt PAYT in response to citizens' demand for, e.g., better service quality, lower waste charges, and fairer waste pricing \citep{Dijkgraaf2009, Batllevell2008}. Public engagement may correlate with waste behaviors as well as with policy acceptance and reactions to policy. To proxy for political participation, I include municipal level voter turnouts in the 2013 Italian general election. Political polarization and lack of social cohesion may also impact policy adoption and responses on waste generation. Protest votes and extreme ideology are proxied by vote shares for big-tent parties, and for extreme left- and right-wing parties.\footnote{Big-tent (or ``catch-all") parties do not rely on strong ideology, and aim to attract voters with diverse political viewpoints, appealing to a large amount of the electorate.}

Moreover, mayor characteristics can matter for policy adoption decisions and also for waste outcomes because they may shape waste-friendly behaviors. Newly elected, young mayors may be willing to invest in waste technology; citizens may elect green mayors putting PAYT in their agenda; locally born mayors of regional parties may promote pollution-saving behaviors, and may rely on large consensus and push reforms. To control for this, I add mayor's age, term length, party, and place of birth.

\vspace{0.5cm}
\textit{Pre-policy waste generation.}--Lagged waste generation may drive policy adoption as well as effect heterogeneity \citep{Kinnaman2006}. In fact, lagged waste outcomes reflect initial opportunity costs of household recycling and waste avoidance, and account for existing differences in the recycling infrastructure, e.g., at the curb. The waste generation history also contains information about unobservables such as motivation and experience in waste reduction, and pro-environmental attitudes \citep{Valente2019}. Including lagged waste amounts, therefore, accounts for possible time-varying effects of fixed unobservables.

\vspace{0.5cm}
\textit{Pre-policy waste management costs.}--Policy adoption decisions largely rely on municipal cost levels, both in per capita and per kg terms. If salient to the household, lagged per capita costs may also impact household waste generation. Cost variables include budgetary costs for waste collection (e.g., labor), disposal (e.g., machinery and land), transportation (e.g., trucks), treatment, and administrative services, net of recycling revenues from selling products and energy recovery.

\section{Modeling Framework} \label{Sec:theorymethod}

I am interested in how a price on household unsorted waste disposal affects waste generation and costs at municipal level. Municipal costs are measured as the sum of waste management costs of unsorted and recycling waste. How do PAYT-induced changes in household waste generation affect municipal costs? This depends on the relative size of cost savings from unsorted waste reductions versus additional costs of increased recycling. Thus, empirically, it is key to understand the behavioral mechanism behind PAYT causal effects, namely, the extent to which unsorted waste reductions translate into more recycling and lower total waste.

To guide the interpretation of my empirical findings presented below, I construct a simple model of household behavior building on the seminal contribution of \cite{Kinnaman17book}. The model delivers two key empirical predictions. 
First, after PAYT introduction, households are predicted to increase recycling and reduce total waste. Second, household responses are heterogeneous depending on, in particular, income and waste generation habits.

\subsection{Theoretical model} \label{Sec:motivation_method}
Consumption generates waste. A household generates unsorted waste $UW$ from consumption, and spends time and effort to recycle the amount $RW$. The opportunity cost of time is the hourly wage $w$, and the marginal cost of recycling is $\chi RW$ where $\chi$ is the amount of effort required for recycling.  Proxies for $\chi$ are household demographics and other waste generation determinants described in Section 1.4. Households can work $H$ hours for a salary $wH$ and can spend part of this time  $h_{RW}$ in sorting recyclable materials. Total waste $TW=RW+UW$ is a fraction $1/\alpha$ ($<$1) of the composite consumption good $c$, such that $TW=(1/\alpha)c$ or $c=\alpha TW$. Households pay a price $p_c$ for consumption and $p_{UW}$ for unsorted waste disposal. The price $p_{UW}$ is the PAYT price, if any, charged by the municipality for each unit of $UW$.\footnote{More generally, the cost of UW generation also includes external costs in terms of, e.g., aesthetic and health costs \citep{Kinnaman1995}. Analyzing these costs is, however, outside the
scope of this paper.} The household utility-maximizing problem is:
\begin{equation} \label{eq1ut}
  U=U(c) \ \ \ \text{subject to} \ \ \ wH-wh_{RW}=p_{c}c+p_{UW}UW,
\end{equation}
where $U(c)$ is utility from consumption ($U_c>0$, $U_{cc}<0$). Equation (\ref{eq1ut}) states that households maximize utility from consumption and use their income $wH-wh_{RW}$ to pay for consumption and unsorted waste disposal. Substituting for $c=\alpha TW$ and assuming the recycling cost function $h_{RW}=h(RW)=\frac{\chi RW^2}{2}$ as in \cite{Kinnaman17book} leads to the Lagrangian:
 \begin{equation} \label{eq2lag}
 \mathcal{L}=U[\alpha(RW+UW)]+\lambda[wH-w(\frac{\chi RW^2}{2})-p_{c}\alpha(RW+UW)-p_{UW}UW],
 \end{equation}
where $\lambda$ is the marginal utility of income. Solving for the first-order conditions (reported in Appendix \ref{Sec:appendix0}) shows that $p_{UW} = w\chi RW$, namely, households choose the optimal amount of recycling when the marginal cost of recycling equals the marginal cost of unsorted waste disposal.

How does an exogenous increase in $p_{UW}$  due to PAYT adoption affect the optimal choice of $RW^*$ and $UW^*$? Using the first-order conditions, I derive three key comparative statics that will form the basis for my empirical analysis. The proofs are relegated to Appendix \ref{Sec:appendix0}.%\footnote{Proofs in Appendix consists in solving the first-order conditions of $RW^*$ and $UW^*$, and then differentiate that with respect to $p_{UW}$.}

\vspace{0.3cm}
\textit{Result 1.-- After PAYT introduction, households reduce unsorted waste, an effect driven by an increase in recycling and a reduction in total waste:}
 \begin{gather} \label{cs1}
\frac{\delta UW^*}{\delta p_{UW}}= -\frac{1}{w\chi} -\frac{UW^*}{\alpha U_c/\lambda^*}<0 \\ \label{cs2}
 \frac{\delta RW^*}{\delta p_{UW}}=\frac{1}{w\chi} > 0.
\end{gather}

This result shows that households decrease unsorted waste due to two effects. Equation (4) shows an increase in recycling (``substitution" effect). Equation (5) shows a reduction in total waste due to less consumption (``income" effect):
 \begin{equation} \label{cs3}
     \frac{\delta TW^*}{\delta p_{UW}}=\frac{\delta RW^*}{\delta p_{UW}}+\frac{\delta UW^*}{\delta p_{UW}} ={\frac{\delta RW^*}{\delta p_{UW}}}-{\frac{\delta RW^*}{\delta p_{UW}}}-\frac{UW^*}{\alpha U_c/\lambda^*}=-\frac{UW^*}{\alpha U_c/\lambda^*}<0.
 \end{equation}

\vspace{0.2cm}
\textit{Result 2.-- The effects of PAYT are heterogeneous depending on, in particular, income and waste generation habits.}

\vspace{0.2cm}
First, the increase in recycling is larger for households with lower costs of time and effort, namely, low-wage households (low $w$) and households who find recycling easier (low $\chi$). Second, the reduction in total waste is larger for high-waste households (high $UW^*$, low $\alpha$), and for households with higher marginal utility of income (high $\lambda^*$) and lower marginal utility of consumption (low $U_c$). Therefore, I expect that, especially, income and pre-policy levels of recycling and unsorted waste can be important predictors of PAYT effect heterogeneity. 

Summing up, this simple model predicts that, after PAYT adoption, municipalities experience a reduction in unsorted waste, an increase recycling and a decrease in total waste. The heterogeneity in household responses, coupled with the large heterogeneity in municipal waste management costs (see Table \ref{tab:nevertreated}), make the sign and magnitude of PAYT effects on municipal costs ex-ante ambiguous and, thus, an empirical question.

\subsection{Identification and empirical specification} \label{Sec:method}

I follow the potential outcome approach \citep{Rubin1974}. Let $(X_{it}, Y_{it},P_{it})$ be the available data for municipality $i = 1,...,n$ at time $t=2010,...,2015$, where $X_{it} \in \R^d$ is a vector of $d$ covariates, $Y_{it}$ is the waste outcome for either unsorted (UW), recycling (RW), or total waste (TW), and $P_{it} \in  \mathcal{P}=[0;p_{max}]$ is the PAYT price (fee) or ``treatment" variable in year $t$. Henceforth, I omit the true subscript $t$ for simplicity whenever possible. Note that prices range between zero (for all untreated years, namely, pre-treatment years for PAYT municipalities or any year for non-PAYT municipalities) and the maximum price set in treated years ($p_{max}$). For every unit $i$, there is a set of potential waste outcomes $Y_i(p)$, $p \in \mathcal{P}$, each being a random variable mapping a particular potential treatment, $p$, onto a potential outcome such that $Y_i=Y_i(p)$. This is also referred to as the unit level dose-response function. For any municipality defined by a vector of characteristics $X_i=x$, I wish to estimate the individual treatment effect of unit $i$, which is defined as $Y_i(p) - Y_i(0)$ and is, however, unobserved for any unit. Therefore, I will estimate the Conditional Average Treatment Effect (CATE) function and its derivative, the Conditional Average Price Effect (CAPE) function:
\begin{gather} \label{eq:catecape}
CATE=   \Delta(x) = \e{Y_i(p) - Y_i(0)|X_i=x}\\
CAPE=   \delta(x)= \pdv{\e{Y_i(p)|X_i=x}}{p},
\end{gather}
under the ``canonical" assumptions of unconfoundedness and no spillovers (more precisely, the SUTVA or Stable Unit Treatment Value Assumption as in \citealt{RosenRubin83, ImbensRubin2015}). I refer to Hirano and Imbens (\citeyear{Hirano2004}) and Imai and van Dyk (\citeyear{Imai2004}) for a description of these assumptions. In essence, unconfoundedness requires to have enough controls - usually pre-treatment covariates and outcomes - so that, conditional on those controls, treatment assignment is as good as randomized. In the case of multivalued treatment, this assumption writes $Y_i(p) \independent P_i | X_i$ $\forall p \in [0;p_{max}]$, i.e., requires conditional independence to hold for each value of the treatment. \citet{Imbens2000} referred to this as weak unconfoundedness, since it does not require joint independence of all potential outcomes.\footnote{Yet, it is difficult to think of applications where the weaker form would be plausible but the stronger form would not be. Differences between the two are
rather conceptual \citep[see][for details]{Imbens2000}.} While this assumption is not directly testable, I make it plausible by controlling for a large set of covariates and pre-policy outcomes. I further assess unconfoundedness in Section \ref{sec:robustness1}.

SUTVA excludes the possibility of interference between units and, given the observed covariates, allows to consider the potential outcomes of one unit to be independent of another unit's treatment status. To make this assumption more reliable, I control for information effects across treated units, and I assess robustness to spillovers in untreated units (see Sections \ref{Sec:sel} and \ref{sec:robustness1}).

In order to fulfill unconfoundedness in the case of non-random price adoption, one needs to control for the sources of self-selection, i.e., capture the effect of $X_i$ on $P_i$. Consider the partially linear waste outcome model:
\begin{equation}\label{eqY}
Y_i = \e{Y_i(0)|X_i} + P_i\delta(X_i) + \epsilon_i(P_i).
\end{equation}
Under unconfoudedness, $\e{\epsilon_i(P_i)|X_i,P_i}=0$; further, $\e{Y_i(0)|X_i}$ represents the possibly non-linear direct effect of covariates on untreated outcomes. Note that $X_i$ is of potentially very high dimension. Let $\e{Y_i | X_i}$ be the conditional outcome mean, and $\e{P_i | X_i}$ the conditional price mean (or ``propensity score"), respectively. By means of algebraic transformations (see proofs in Appendix \ref{Sec:appendix00}), model (\ref{eqY}) can be rewritten as: 
\begin{equation}\label{eqYres}
 Y_i - \e{Y_i | X_i} = (P_i - \e{P_i | X_i})\delta(X_i) + \epsilon_i(P_i),
\end{equation}
where $\delta(X_i)$ identifies the CAPE as the effect of the leftover price variation $P_i - \e{P_i | X_i}$ on the leftover outcome variation $Y_i - \e{Y_i | X_i}$ not explained by the observed covariates. The standard partially linear model considers solely the case of constant treatment effects  \citep{Robinson1988}. Yet, \cite{nie2019} study identification of model (\ref{eqYres})   for flexible heterogeneous treatment effect estimation via machine learning approaches. In particular, we can estimate $\delta(x)$ in two steps: First, we separately estimate the nuisance components $\e{Y_{i} | X_{i}=x}$ and $\e{P_{i} | X_{i}=x}$. Second, we plug in their fitted values to obtain $\hat{\delta}(x)$ by regressing residualized outcomes on residualized prices. In this step, we do not consider all residuals as equally important, and we estimate a local version of model (\ref{eqYres}) that gives more weight to those residuals in the neighborhood of $x$ (see the next subsection for details).

Residual-on-residual regression methods, also known as R(esidualized)-learning in high-dimensional settings, make the parameter estimate insensitive to small errors in the nuisance components, thus improving its robustness. The final estimator combining predictions from two models is robust to possible misspecifications in either the outcome model or the propensity score model \citep{nie2019}. Note that, as an alternative approach, one could directly estimate model (\ref{eqY}) by including the high-dimensional covariate set. However, regularization of $\e{Y_i(0)|X_i}$ would cause an especially large bias when covariates are correlated with prices \citep{athey2017, Chernozhukovetal2017}.\footnote{Without regularization, the inclusion of a large set of partly correlated predictors may lead to, e.g., variance inflation and incorrect signs.} Conversely, residualization removes the correlation of covariates with both prices and outcomes, rendering the estimator robust to the parametric form in which covariates are included. In particular, as long as either the estimator for either propensity scores or conditional outcome expectation is consistent, the resulting estimator for the treatment effect is consistent \citep{nie2019}.

\subsection{Random forests for heterogeneous policy effects} \label{sec:RF}

I now discuss how to estimate equation (\ref{eqYres}) for each municipality $i$ described by the covariate vector $X_i=x$. First, I estimate $\e{P_i | X_i}$ for $X_i=x$, which defines the generalized propensity score $s(x)\coloneqq\e{P_i | X_i=x}$. Second, I estimate $\e{Y_i | X_i}$ for $X_i=x$, which defines the expected outcome marginalizing over treatment $y(x) \coloneqq \e{Y_i | X_i=x}$. Third, I use $s(x)$ and $y(x)$ to estimate $\delta(X_i)$ for $X_i=x$, which defines the CAPE $\delta(x)$. To explicitly account for heterogeneity in the parameter estimation procedure, I propose to use the Random Forest (RF) estimator developed in \citet{atheyGRF}. This method generalizes the original algorithm of \citet{Breiman2001} by adapting to the problem of heterogeneous treatment effect estimation. 

%the unit level predictions for the conditional expectations of interest: as mentioned above, I estimate first the generalized propensity score $s(x)\coloneqq\e{P_i | X_i=x}$ and the expected outcome marginalizing over treatment $y(x) \coloneqq \e{Y_i | X_i=x}$, and finally the PAYT causal effect $\delta(x)$ or CAPE.

RFs present several advantages compared to other methods. First, standard $k$-nearest neighbor/kernel matching methods are bound to fail, as the concept of neighbor vanishes in high-dimensions \citep{Imbens2006, Giraud2015}. Conversely, RFs are adaptive (data-driven) matching estimators that can estimate neighbors in a high-dimensional space while avoiding overfitting through the use of both training and estimation samples. Second, RFs are nonparametric techniques relying on recursive, data-driven sample splits to capture possibly
complex non-linear interactions between covariates and treatment effects without assumptions regarding the data distribution. This reduces the risk of model misspecification and limits researcher discretion when selecting the relevant dimensions of heterogeneity. Third, and most importantly, RFs are consistent estimators of the full mapping of CAPE; in other words, they allow to construct a policy targeting function mapping observed covariates onto unit level causal effects. 
 
 Breiman's RF is understood as an ensemble method averaging predictions made by individual trees. Yet, as shown in \cite{atheyGRF}, we can equivalently think of RF as an adaptive kernel method. For instance, we can re-write the forest prediction as $\hat{y}(x)=\sum_{i=1}^n w_i(x)Y_i$ where $w_i(x)$ is a data-adaptive kernel that measures how often the $i^{th}$ training unit falls into the same leaf as $x$. How does the RF determine the weighted set of neighbors for any test point $x$?

 \vspace{0.3cm}
   \textit{Step 1:} \textit{Random subsampling.} Grow a
set of $B$ trees and, for each tree, $b=1,\dots,B$, draw a random subsample of training data $s_b^{tr}\subseteq\{1,\dots,n\}$.

 \vspace{0.3cm}
   \textit{Step 2:} \textit{Recursive partitioning.}   Estimate a tree via recursive partitioning on $s_b^{tr}$ and a random subset of covariates. During each partitioning step, split the training sample according to any cutpoint of each covariate, and obtain two new partitions. In each partition, run model (\ref{eqYres}) to estimate $\hat{\delta}(x)$ and its variance across partitions. Repeat this step using every covariate and select the one that leads to the largest CAPE heterogeneity (variance). Split the data into two subgroups based on the optimal cutpoint in the selected covariate. Thus, the selected predictor becomes a partitioning variable. Continue partitioning $s_b^{tr}$ a number of times until the splitting process terminates after a particular stopping criterion is reached.  The resulting tree is a sequence of binary regions partitioning the covariate space and grouping observations in the bottom regions called leaves $L_b$.

 \vspace{0.3cm}
   \textit{Step 3:} \textit{Prediction.}  Define $L_b(x)$ as the set
of training observations falling in the same neighborhood (leaf) as $x$. Each tree gives equal (positive) weight to the observations in the
same leaf as the test point $x$, and zero weight to all the other observations. Then, the forest averages all these tree-based weightings, measuring how often each training example $i$ falls into the same leaf as $x$. This measure is given by the forest weights $w_i(x)$: 

\begin{equation}
    w_{ib}(x)= \frac{\mathbbm{1}({X_i \in L_b(x)})}{|L_b(x)|} \ \ , \ \ w_i(x)= \frac{1}{B} \sum_{b=1}^B  w_{ib}(x),
\end{equation}
where $\sum_{i=1}^n w_i(x)=1$. For more details, see \cite{atheyGRF}.

 \vspace{0.3cm}
This kernel-based perspective of RFs allows to cast this method as an adaptive locally weighted estimator that first uses a forest to calculate a weighted set of neighbors for each test
point $x$, and then solves a weighted version of the residual-on-residual regression (\ref{eqYres}). The resulting CAPE estimator writes:
\begin{equation}\label{eq:rlearner}
    \hat{\delta}(x)= \frac{\sum_{i=1}^n w_i(x)(Y_i-\hat{y}^{-i}(X_i))(P_i-\hat{s}^{-i}(X_i))}{\sum_{i=1}^n w_i(x)(P_i-\hat{s}^{-i}(X_i))^2},
\end{equation}
where $\hat{s}^{-i}(X_i)$ are out-of-sample predictions of the propensity scores $s(x)$. Equation (\ref{eq:rlearner}) characterizes the CAPE as a weighted residual-on-residual regression estimator with forest weights $w(x)$ defining the nearest neighbors of $x$ produced by different trees.\footnote{RFs have tuning parameters such as minimum leaf-size and penalties for imbalanced partitions. As will be discussed later (see Section \ref{Sec:results}), these are obtained via cross-validation as in \cite{grfpack}, i.e., choosing the ones that make out-of-sample estimates of the regression errors minimized in the R-learning objective as small as possible.} This distinguishes the CAPE from the APE (Average Price Effect) estimator which is the unweighted version of (\ref{eq:rlearner}).\footnote{Motivated by the R-learning equation (\ref{eqYres}), I estimate the APE as $\frac{\sum_{i=1}^n(Y_i-\hat{y}^{-i}(X_i))(P_i-\hat{s}^{-i}(X_i))}{\sum_{i=1}^n(P_i-\hat{s}^{-i}(X_i))^2}$.} Variance of $\hat{\delta}(x)$ is estimated by evaluating the estimator on bootstrapped half-samples of the training data, also called bootstrap of little bags \citep{atheyGRF, laake2009}.

%%%%%%%%%%%%%%%%%%%%% RESULTS
\section{Main Results} \label{Sec:results}

I estimate separate random forests for each outcome (unsorted, recycling, total waste per capita) and policy year.\footnote{Policy years range between one and three as treated units are observed up to three years after policy implementation. Note that waste demands can be estimated separately although different waste types are determined simultaneously. This does not generate bias because waste demands are functions of the same variables \citep{Fullerton2000}. Reassuringly, the adding up condition also holds after estimation, namely, differences between $\widehat{UW}+\widehat{RW}$ and $\widehat{TW}$ are negligible.} A tree is built with data for the same calendar year to account for common shocks to all units. In order to account for the staggered nature of the intervention, treated units in a given policy year are matched with never-treated units only. In other terms, late adopters are not included in the control group of early adopters. This is important especially since lagged outcomes are included among the covariates. 

I have explored from 500 to 10,000 trees in the RF, and treatment effect estimates become stable after 1,000 trees, thus, results are obtained using this value. All trees are grown with cross-validated values for the number of randomly subsampled covariates, minimum leaf size, and penalty for imbalanced splits, namely, splits in which the size of parent and child node are very different are penalized. To overcome the issue that the treatment group is substantially smaller than the control group, each node is required to include a minimum number of both treated and control units, i.e., enough information about both factual and counterfactual to estimate the treatment effect reliably. For this reason, a penalization is imposed also to nodes including an unbalanced number of treated and control units. Following \citet{atheyGRF}, values for such parameters are obtained via cross-validation.\footnote{For the entire RF analysis, I use the software R-3.4.2 and the grf package version 0.10.0 \citep{grfpack}.} 

In order to credibly estimate policy causal effects, the common support condition for treated and untreated units needs to hold. Figure \ref{fig:overlap} in Appendix \ref{Sec:appendixB} provides evidence of such support.

\subsection{Policy effects on waste demands}

I begin my empirical analysis by estimating the CAPE function in equation (\ref{eq:rlearner}), namely, I estimate municipal level causal effects of prices on demands for unsorted,  recycling, and total waste per capita (UW, RW, TW, respectively). In this section, I focus the attention on causal effects in the third policy year in order to have a closer estimate to the long-term effects of PAYT.\footnote{I analyze the dynamics of price effects in Section \ref{sec:welf}. One caveat is that these elasticities cannot be estimated for late adopters. Reassuringly, there are no statistically significant differences in the price effect distribution of early and late adopters in the first two policy years.} I run separate random forests for each waste outcome: UW, RW, and TW (for robustness). Figure \ref{fig:main0} presents the CAPE distribution for all municipalities in the sample. Price effects are estimated as price semi-elasticities of waste demands. All estimates are statistically different from zero (p-values $<$ 0.01). The hypothesis of effect homogeneity is rejected for all outcomes and policy years.\footnote{See Levene's tests (\citeyear{Levene1960}) and heuristics \citep{athey2017} in Appendix \ref{Sec:appendixC:het}.} 

\begin{center}
    \begin{minipage}[t]{\linewidth}
      \begin{figure}[H]
   \centering
\subfloat{\includegraphics[scale=0.4]{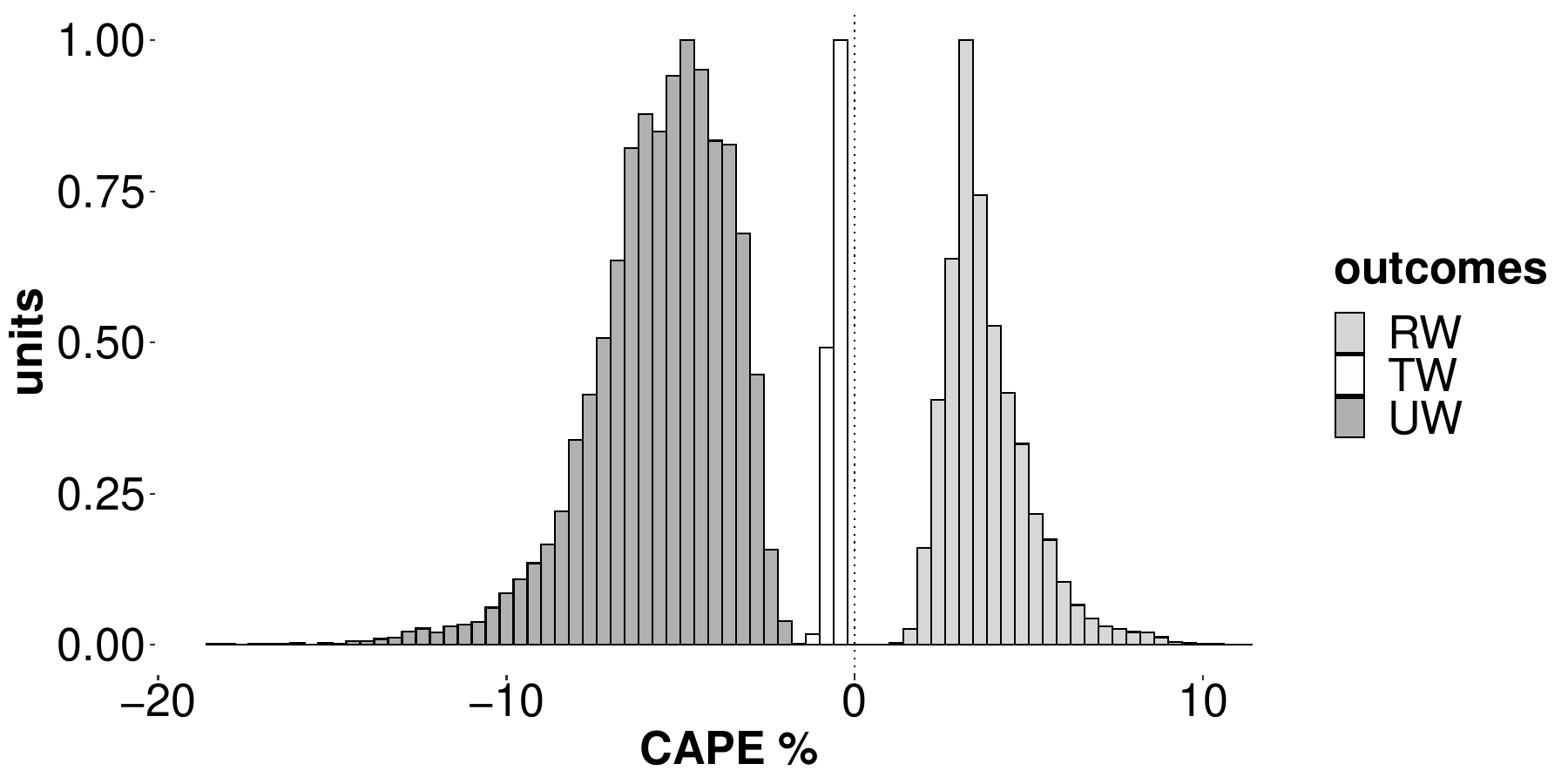}}
   \centering
       \caption{ \footnotesize Unit level estimates of price semi-elasticities (CAPE) of waste demands. CAPE are measured as percent changes of waste amounts for a one cent price increase.}
 \footnotesize
 \label{fig:main0}
    \end{figure}
  \end{minipage}
  \end{center}
 \vspace{0.5cm}
\noindent 
The estimated price effects on waste demands for UW and RW are large in magnitude. I use the estimated CAPE to compute factual and counterfactual waste amounts, and the percent (causal) change in waste relative to the counterfactual. I find that prices cause an annual UW decline of 50\% on average. This reduction is mainly driven by an average RW increase of 32\% and, to a lesser extent, by an average TW reduction of 5\%.\footnote{This corresponds to -110 kg (UW), +80 kg (RW) and -25 kg (TW) per capita. Thereby, estimates from separate forests are consistent with each other ($CATE_{TW} \approx CATE_{UW}+CATE_{RW}$). Also note that the low variance of the effects on TW in Figure \ref{fig:main0} is due to the high and negative correlation between the effects on UW and RW.} The large substitution effect of UW with RW caused by the policy, therefore, represents a strong change in waste behaviors.

\subsection{Sources of causal effect heterogeneity}

Using the CAPE function defined in equation (\ref{eq:rlearner}) and estimated in the previous section, I proceed with the analysis of the sources of effect heterogeneity. In order to summarize the CAPE function, I linearly project the CAPE estimates onto price levels, a set of municipal characteristics, and their interactions. This corresponds to fitting the linear model:
\begin{equation}
    \beta = \argmin_\beta \{\e{(CAPE-X_{sub}\beta)^2}\}
\end{equation}
\noindent
where $CAPE$ are point estimates of the price causal effects (as presented in Figure \ref{fig:main0}), and $X_{sub}$ are explanatory variables including price levels, a subset of municipal characteristics, and their interactions.\footnote{This regression does not account for the uncertainty of the CAPE point estimates. In Appendix \ref{app:capekg}, I show that accounting for uncertainty bounds  of the point estimates does not substantially alter the shape and magnitude of the fitted heterogeneity.} The literature on waste prices is especially interested in learning the elasticity of demand as a function of a few variables such as income or education. Thereby, the set of features includes relevant regressors capturing household opportunity costs of waste disposal: income, education, and pre-policy waste levels. Figure \ref{fig:main1} plots fitted semi-elasticities (CAPE) across prices, ceteris paribus.\footnote{Data inspection suggests quadratic demand curves. Thus, I fit a polynomial regression of order 2. Results for CAPE in kg are presented in Figure \ref{fig:mainkg} in Appendix \ref{app:capekg}.} Pointwise estimates are obtained by fitting a smooth regression line. Shaded regions are the 95\% confidence intervals. 

\begin{center}
    \begin{minipage}[t]{\linewidth}
      \begin{figure}[H]
   \centering
\subfloat{\includegraphics[scale=0.4]{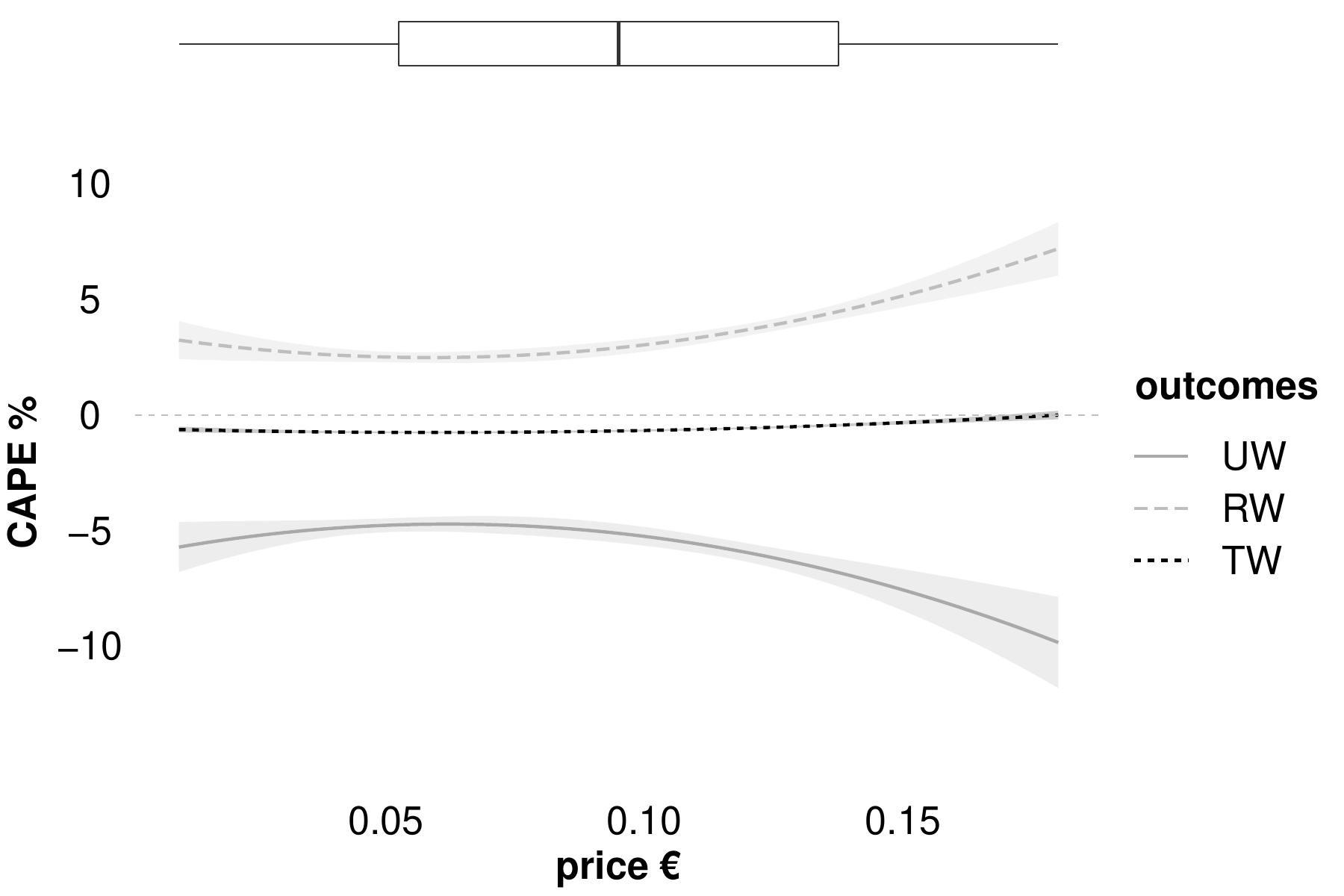}}
   \centering
       \caption{ \footnotesize Fitted price semi-elasticities (CAPE) of waste demands at each price level.}
 \footnotesize
 \label{fig:main1}
    \end{figure}
  \end{minipage}
  \end{center}
 \vspace{0.5cm}
\noindent
Price elasticities of waste demands in high and low price regions differ. I use the median value of 9 cents to define these regions, indicated by the vertical line in the boxplot above the image. While at high prices elasticities are increasing, elasticities are rather constant at low prices. All other things equal, a one cent price increase reduces unsorted waste by 5 to 10\% (vs. 4.7-5.7\% at low prices), increases recycling by 2 to 6\% (2.5-3.2\%) and reduces total waste by 0.1 to 0.7\% (0.6-0.8\%) compared to a counterfactual zero-price (no PAYT) scenario. Further, demand responses at high prices shed new light on the mechanism behind UW reductions: increasing UW reductions reveal a negative correlation between TW and RW reactions, suggesting that waste reduction and recycling are substitute behaviors.\footnote{Section \ref{sec:welfsim} provides estimates of the elasticity of substitution between recycling and waste avoidance behaviors.} Thus, higher prices do not cause further total waste reductions, instead they simply reallocate waste to the recycling pile.

Having established effect heterogeneity across prices, I analyze possible income effects. Figure \ref{fig:mainINCOME} plots fitted semi-elasticities by income level, ceteris paribus.

\begin{center}
    \begin{minipage}[t]{\linewidth}
      \begin{figure}[H]
   \centering
\subfloat{\includegraphics[scale=0.48]{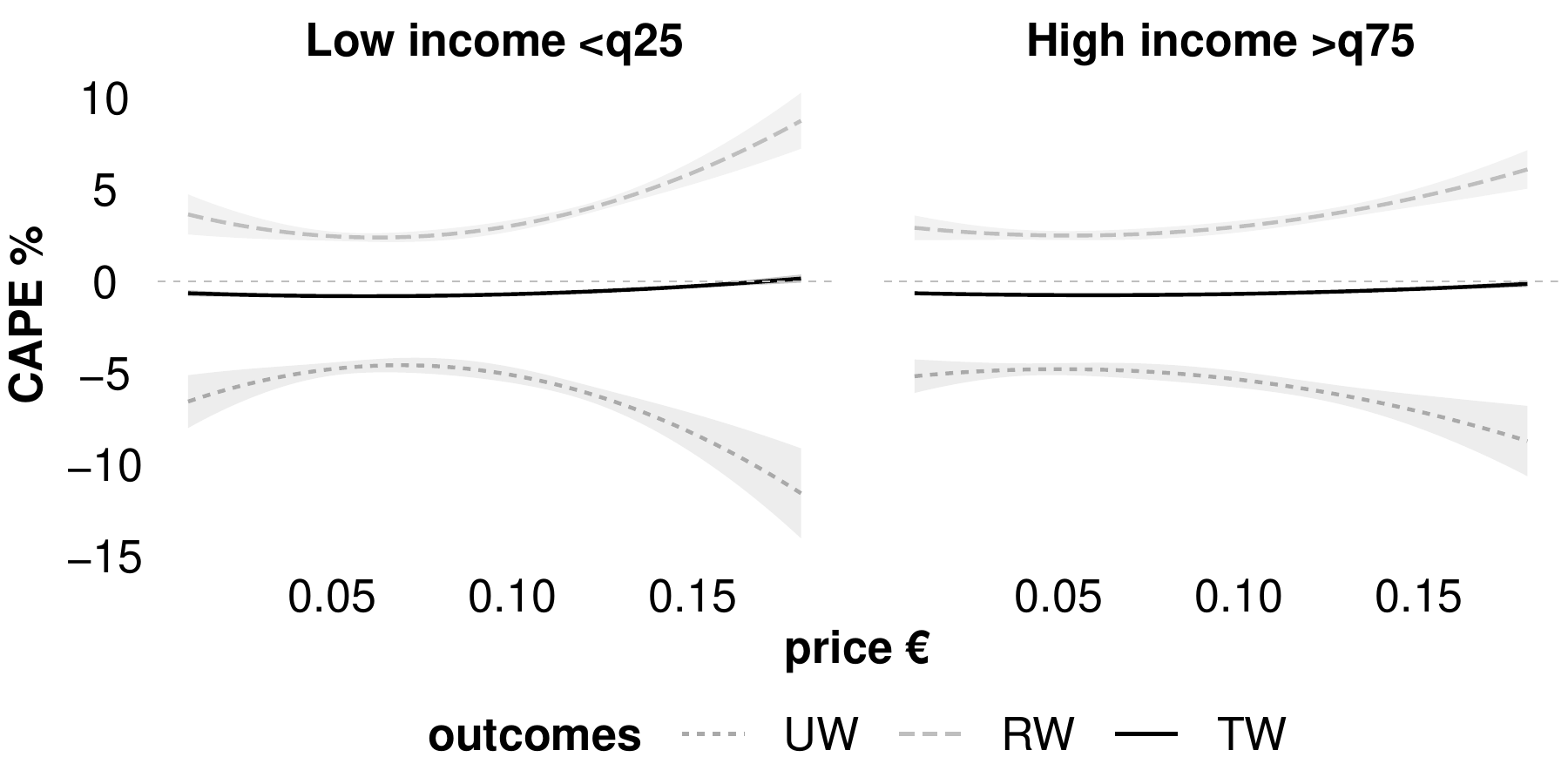}}
      \centering
    \caption{ \footnotesize Fitted price semi-elasticities (CAPE) of waste demands by income levels. Thresholds q25 and q75 indicate first (\EUR{}13k) and third (\EUR{}16k) quartiles of annual income per capita.}
 \footnotesize
 \label{fig:mainINCOME}
    \end{figure}
  \end{minipage}
  \end{center}
 \vspace{0.5cm}
\noindent
Income differences across municipalities significantly explain price effect heterogeneity at high prices: Low-income municipalities are on average one percent point more elastic than high-income municipalities.\footnote{In particular, at high prices, semi-elasticities for low-income municipalities are on average -7.5\% for UW, 5.2\% for RW, and -0.5\% for TW, ceteris paribus.} This result is new to the literature as previous studies typically consider the effects of low prices.\footnote{The highest price in, e.g., \cite{Fullerton2000} is \$2.18 per 32-gallon bag (120L) or \$0.02/L. This data include prices up to \EUR{}0.18/L. See \citet{Bel2014} for a review of previous studies.} Differently, income effects are not statistically significant at low prices. This may be explained by the low monetary benefits of reducing UW at low prices: households reduce UW by 13.6\% in the first price quartile (3 cents) and save less than one euro.

Additionally, price effects at high prices are heterogeneous also in recycling habits before policy, as predicted by \cite{Kinnaman2006}. In particular, high prices are effective in increasing recycling levels in low-recycling municipalities which have large margin to improve on this behavior.\footnote{See Figure \ref{fig:mainRW} in Appendix \ref{app:capekg}.} 

Finally, looking at heterogeneity across education levels, I estimate no significant effects, suggesting that education plays, if any, a relatively minor role.

\subsection{Comparison with the literature}

Typically, the literature estimating demand responses of waste prices relies on the assumption that the mean price and waste quantity are on one point along the linear demand curve, and finds low average price elasticities \citep[see, e.g.,][]{Kinnaman2006, Fullerton2000}. However, prior work analyzes waste programs with low adopted prices.\footnote{For instance, \cite{Callan2006} use a mean price of \$0.012 per gallon which translates into less than one cent per liter.} The average point-price elasticity of unsorted waste demand across 72 studies amounts to -0.34 \citep{Bel2014}. For comparison, I compute the point-price elasticity at mean price and quantity in the first and fourth quartile of the price distribution. I find a value of -0.26 in the lowest quartile, and a value of -2.2 in the highest quartile. In line with the literature, this study consistently estimates low price elasticities of waste demands when prices are low. Additionally, I show that waste demands are significantly more elastic at high prices.

\subsection{Policy effects on municipal costs} \label{sec:welf}
Heterogeneity of policy effects on municipal budgets is driven by both demand and cost effects. For each municipality $i=1,...,n$ with attributes $X_i=x$, I define the policy Effects on Municipal Costs ${EMC}^{P>0}(x)$ in per capita euros as waste management cost savings caused by PAYT ($P>0$):
\begin{equation} \label{eq:SCS2}
{EMC}^{P>0}(x) = - \Bigg[ {CATE}{_{UW}}PC_{UW}^{P>0}(x)+ {CATE}{_{RW}}PC_{RW}^{P>0}(x)
\Bigg],
\end{equation}
where ${CATE} = p\hat{\delta}(x)$ (in kg per capita) and $\hat{\delta}(x)$ are \textit{statistically significant} CAPE point estimates for each municipality and waste type.\footnote{As CAPE is the first derivative of the CATE, then an estimate of the CATE is just the integral of the estimated CAPE, i.e. of  $\hat{\delta}(x)$ (Torricelli-Barrow theorem).} Further, $PC^{P>0}(x)$ are municipal (private) unit costs of unsorted or recycling waste management under PAYT (in \EUR{}/kg). $PC^{P>0}(x)$ are observed for treated units (under treatment) but unobserved for untreated units. In this section, I estimate whether PAYT caused significant changes in unit costs of treated units using RF.\footnote{I regress all the covariates - including municipal attributes, lagged waste and cost values - and a policy indicator on municipal unit costs, running separate forests for UW and RW.} In the next section, I use the estimated RF function to predict unit costs for untreated units under PAYT (see Section \ref{sec:welfsim} for more details).

In my sample, unit costs of unsorted waste are higher than those of recycling for most municipalities. This implies that unsorted waste reductions that do not translate into an increase in recycling are the driving source of municipal cost savings. In other words, I expect demand responses on total waste to matter for effects on municipal budgets.

I start by documenting that demand effects (CATE) on UW and TW are lowest in the first policy year. This is consistent with economic theory showing that elasticities are typically lower in the short-run because fewer alternatives are available \citep{Usui2009}. UW reductions are largely driven by increased recycling and, to a smaller extent, by total waste reductions. In particular, behavior changes toward total waste are relatively smaller and slower. This is consistent with behavioral economic predictions by e.g. \cite{Cecere2014} who argue that changes in consumption habits and subsequent total waste reductions require more effort and learning than recycling.\footnote{For instance, households need to learn how to reuse or buy products with less packaging.}

Figure \ref{fig:main_catetw} plots the distribution of statistically significant demand effects on TW by policy year. 

\begin{center}
    \begin{minipage}[t]{\linewidth}
      \begin{figure}[H]
   \centering
\subfloat{\includegraphics[scale=0.3]{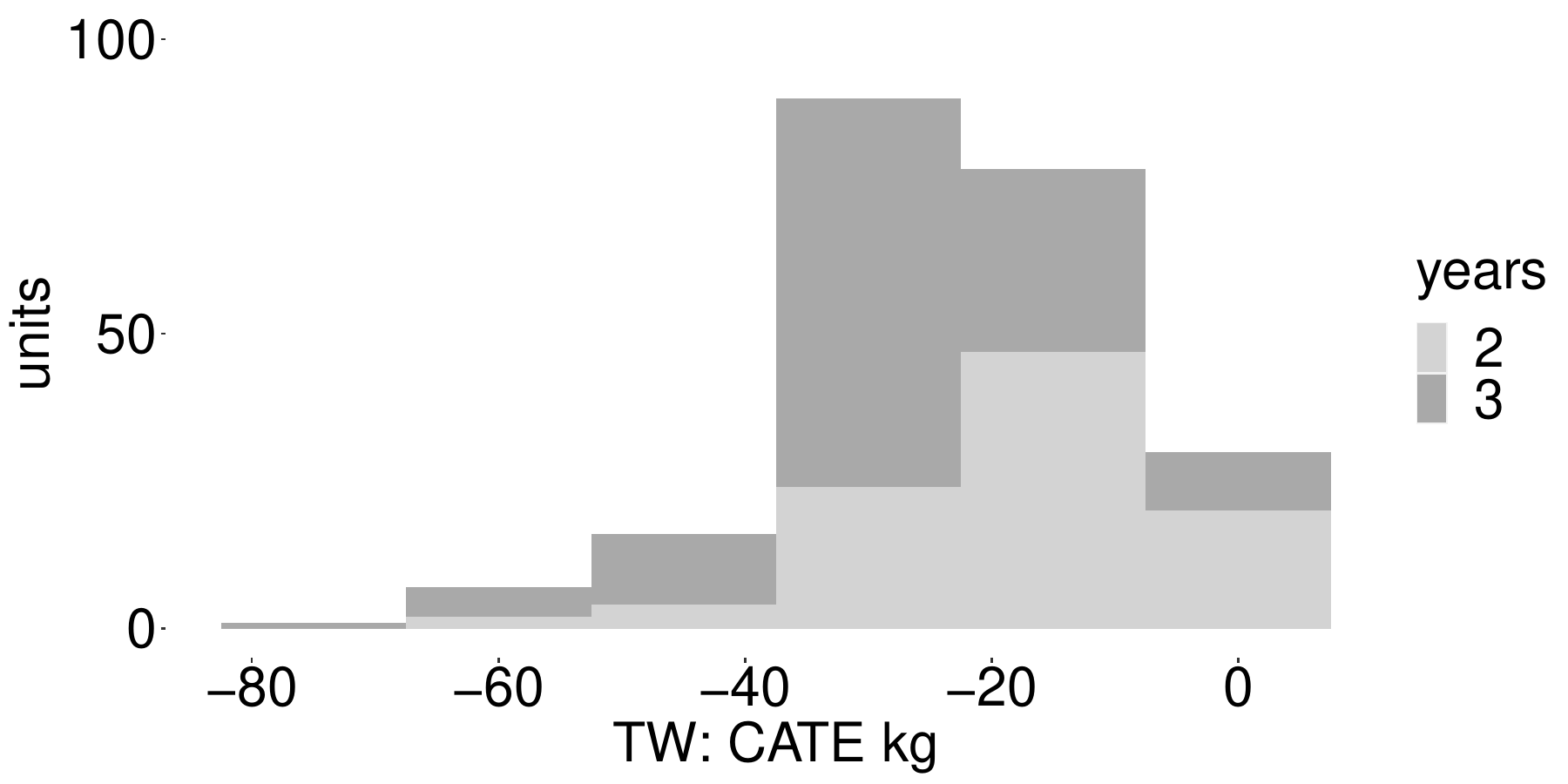}}
       \caption{ \footnotesize Unit level estimates of causal effects (CATE) on total waste demands in the second and third policy year (p-values $<$ 0.05). The y-axis refers to the number of treated municipalities.}
 \footnotesize
 \label{fig:main_catetw}
    \end{figure}
  \end{minipage}
\end{center}
 \vspace{0.5cm}
\noindent
In year 1, there are no statistically significant effects on total waste. In year 2, I observe significant reductions for about 30\% of municipalities, and significant reductions for all municipalities in year 3. 

Next, I flexibly estimate policy causal effects on unit management costs for UW and RW via random forests. I find that PAYT leaves unit costs of waste mostly unaffected. %, suggesting constant returns to scale.%\footnote{This is consistent with results on Italian municipalities by, e.g., \cite{Abrate2014}.}
%Overall, I find no economies of scale, confirming results on Italian municipalities by e.g. \cite{Abrate2014}. In a fraction of municipalities, estimates suggest economies of scale for UW and RW because their unit cost increases (decreases) when UW (RW) reduces (increases). 
Finally, I compute policy effects on municipal costs by combining CATE with unit costs of waste management. Table \ref{tab:welfaremain}  shows municipal cost effects (\EUR{} p.c.) by policy year. 

\renewcommand{\arraystretch}{1.2}

\begin{table}[H]
\centering
\footnotesize
\begin{tabular}{llrrrr}

 &  Year  & Mean & Sd & EMC$>$0\\ 
 \hline
EMC \EUR{} p.c. &1  & 0.13 & 10.16 & 36\% \\ 
 
&2  & 3.14 & 7.10 & 89\%\\

&3   & 20.17 & 16.28 & all \\ 

   \hline
\end{tabular}
 \caption{ \footnotesize Summary statistics for policy Effects on Municipal Costs (EMC) for each PAYT municipality. EMC$>0$ indicate cost savings per capita (p.c.) by policy year.}
 \label{tab:welfaremain}
\end{table}
\noindent
Policy effects on municipal costs are mostly positive. However, waste prices raise municipal costs in a fraction of municipalities where the reduction of unsorted waste is largely driven by an increase in recycling, and management costs of recycling are high compared to those of unsorted waste. In year 3, PAYT generates cost savings in all municipalities. Average savings are \EUR{}20 per capita, which is about one fifth of what municipalities spend for waste management per person on average. As management unit costs are largely unaffected, savings come from UW reductions that do not translate into higher RW. In other words, municipal cost savings are largely driven by total waste reductions. 

As higher prices promote increased recycling rather than waste avoidance, the largest cost savings occur in municipalities setting low prices after three years of adoption.  Figure \ref{fig:main_welfareprices} shows effects on municipal costs at high vs. low prices (above vs. below the median) in the third policy year, when total waste reductions are highest.

\begin{center}

    \begin{minipage}[t]{\linewidth}
      \begin{figure}[H]
   \centering
\subfloat{\includegraphics[scale=0.33]{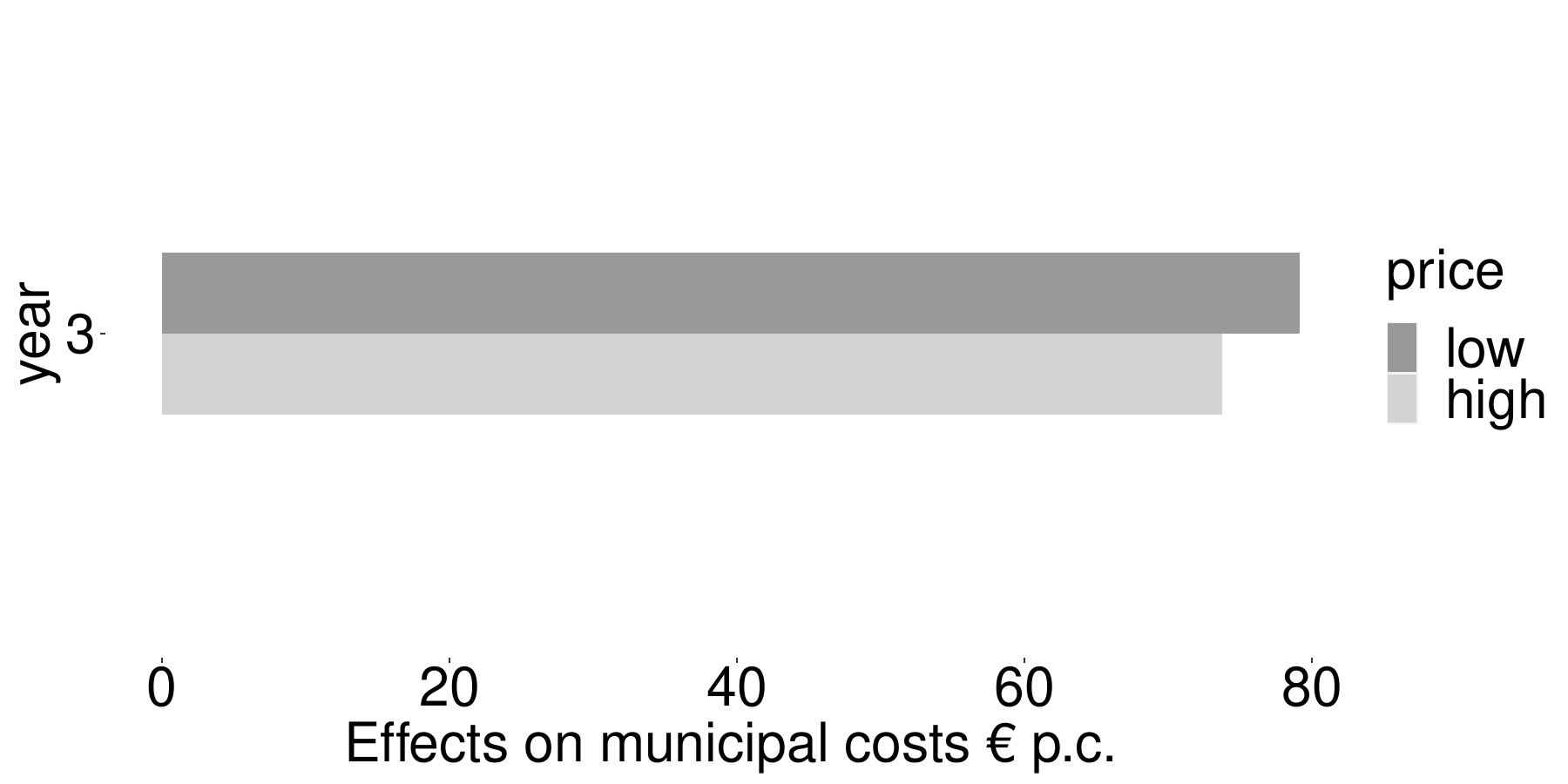}}
       \caption{ \footnotesize Policy Effects on Municipal Costs (EMC$>$0 indicate cost savings) at high vs. low waste prices (above vs. below the median) after three years of adoption.}
 \footnotesize
 \label{fig:main_welfareprices}
    \end{figure}
  \end{minipage}
  \end{center}
 \vspace{0.5cm}
\noindent
This implies that (i) higher prices may be less desirable from a municipal cost perspective as they do not cause further total waste reductions, instead they rather reallocate waste to the recycling pile, and (ii) targeting total waste reductions in high-price municipalities could increase savings substantially.

The final implication of this paper may help local governments to overcome their initial concerns regarding the implementation of PAYT policies.\footnote{In Italy, less than 8\% of municipalities adopt PAYT \citep{ISPRA2019}. Policymakers are concerned about cost increases due to, e.g., higher recycling \citep{facchini2020}. For anecdotal evidence, see e.g. \citet{GilliBook2018}, \citet{Allers2010}, \citet{Callan1999}.} This paper provides evidence that short-term costs may indeed increase due to higher recycling, but these costs are outrun by the long-term savings due to total waste reductions.

\subsection{Policy simulations for municipalities without PAYT} \label{sec:welfsim}

 I estimate causal effects on waste demands and municipal costs if all municipalities in the sample were to implement PAYT. The estimation proceeds in four steps.
 
  \vspace{0.3cm}
 \textit{Step 1: Predict counterfactual prices.} To each non-PAYT municipality, assign the price of the PAYT municipality with the closest predicted propensity score. 
 
  \vspace{0.3cm}
 \textit{Step 2: Predict counterfactual ${CAPE}$.} Predict price semi-elasticities for non-PAYT municipalities using the RF function mapping $X$ onto causal changes in waste amounts previously estimated for PAYT municipalities. 
 
  \vspace{0.3cm}
  \textit{Step 3: Predict counterfactual ${CATE}$.} Predict price effects (in kg) for non-PAYT municipalities multiplying statistically significant CAPE point estimates by the predicted PAYT price. 
  
   \vspace{0.3cm}
   \textit{Step 4: Predict counterfactual ${PC}^{P>0}$.} Predict municipal private unit costs for non-PAYT municipalities using the RF function mapping $X$ onto causal changes in unit costs previously estimated for PAYT municipalities. For each non-PAYT municipality, assess whether PAYT would affect unit costs significantly. If this is the case, replace the observed $PC^{P>0}$ in the absence of PAYT with its predicted counterpart under PAYT.

  \vspace{0.3cm}

Figure \ref{fig:cape_map} plots the predicted price semi-elasticities of RW and TW demands after three years of adoption. Color differences in the same area indicate that higher recycling is associated to smaller reductions in total waste and vice versa.

 \vspace{-0.5cm}
  \begin{minipage}[t]{0.5\linewidth}
    \begin{figure}[H]
 \hspace{-0.7cm}
\subfloat {\includegraphics[trim={1.5cm 0 0 1cm},clip, scale=0.7]{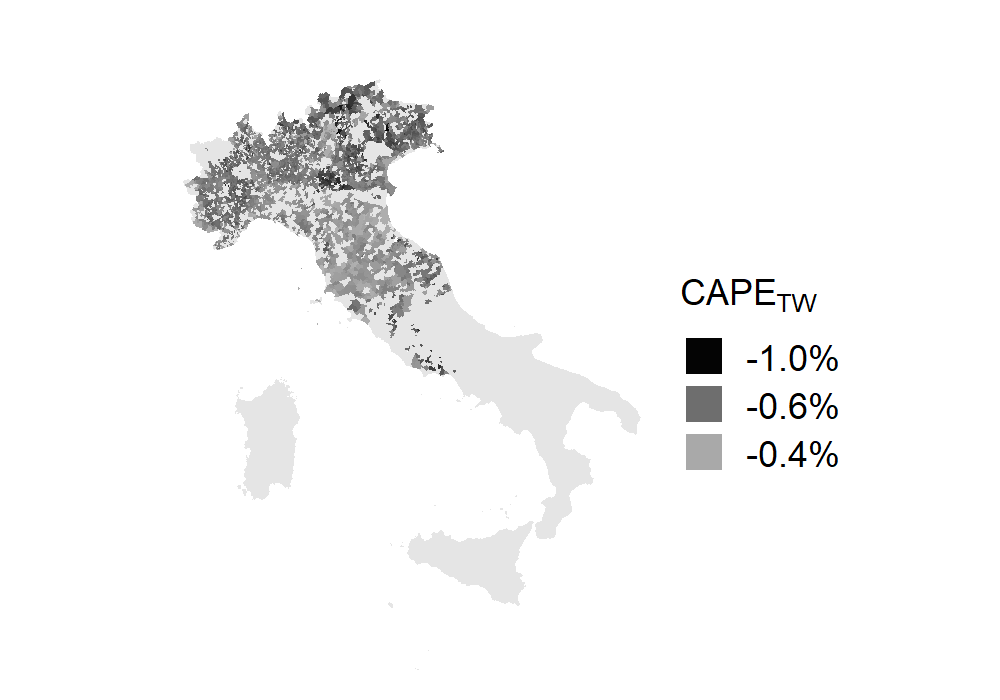}}
 \footnotesize
    \end{figure}
  \end{minipage}  
    \begin{minipage}[t]{0.5\linewidth}
      \begin{figure}[H]
      \hspace{-0.5cm}
\subfloat{\includegraphics[trim={1cm 0 0 1cm},clip, scale=0.7]{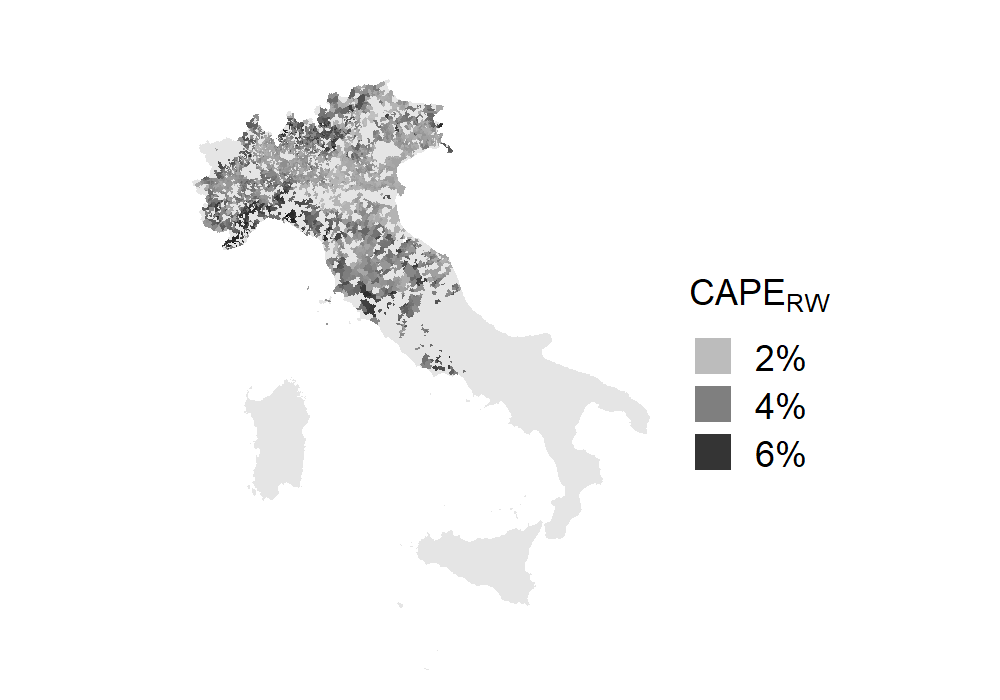}}
 \footnotesize
    \end{figure}
  \end{minipage}
    \captionof{figure}{\footnotesize Predicted semi-elasticities (CAPE) on total waste and recycling demands for all municipalities three years after policy.} \label{fig:cape_map}
  
  \vspace{0.5cm}
\noindent
\setlength\parindent{0.5cm}
In particular, I estimate that a one percent point increase in $CAPE_{RW}$ decreases $CAPE_{TW}$ in absolute terms by on average 0.1 percent points (see Figure \ref{fig:cape5} in Appendix). Importantly, the elasticity of substitution between the effects on recycling and total waste does not lead to rebound effects (i.e., higher total waste).

Despite substitutabilities between waste reduction and recycling behaviors, policy simulations show that most municipalities would benefit from PAYT adoption. Table  \ref{tab:welfarepred} shows predicted policy effects on municipal costs by policy year if all municipalities were to implement PAYT (distributions are plotted in Figure \ref{fig:hist_swe} in Appendix \ref{Sec:appendixD:swe}).

\renewcommand{\arraystretch}{1.2}

\begin{center}
        
      \begin{figure}[H]
   \centering
\subfloat{\includegraphics[trim={0cm 1cm 0 0},clip, scale=0.5]{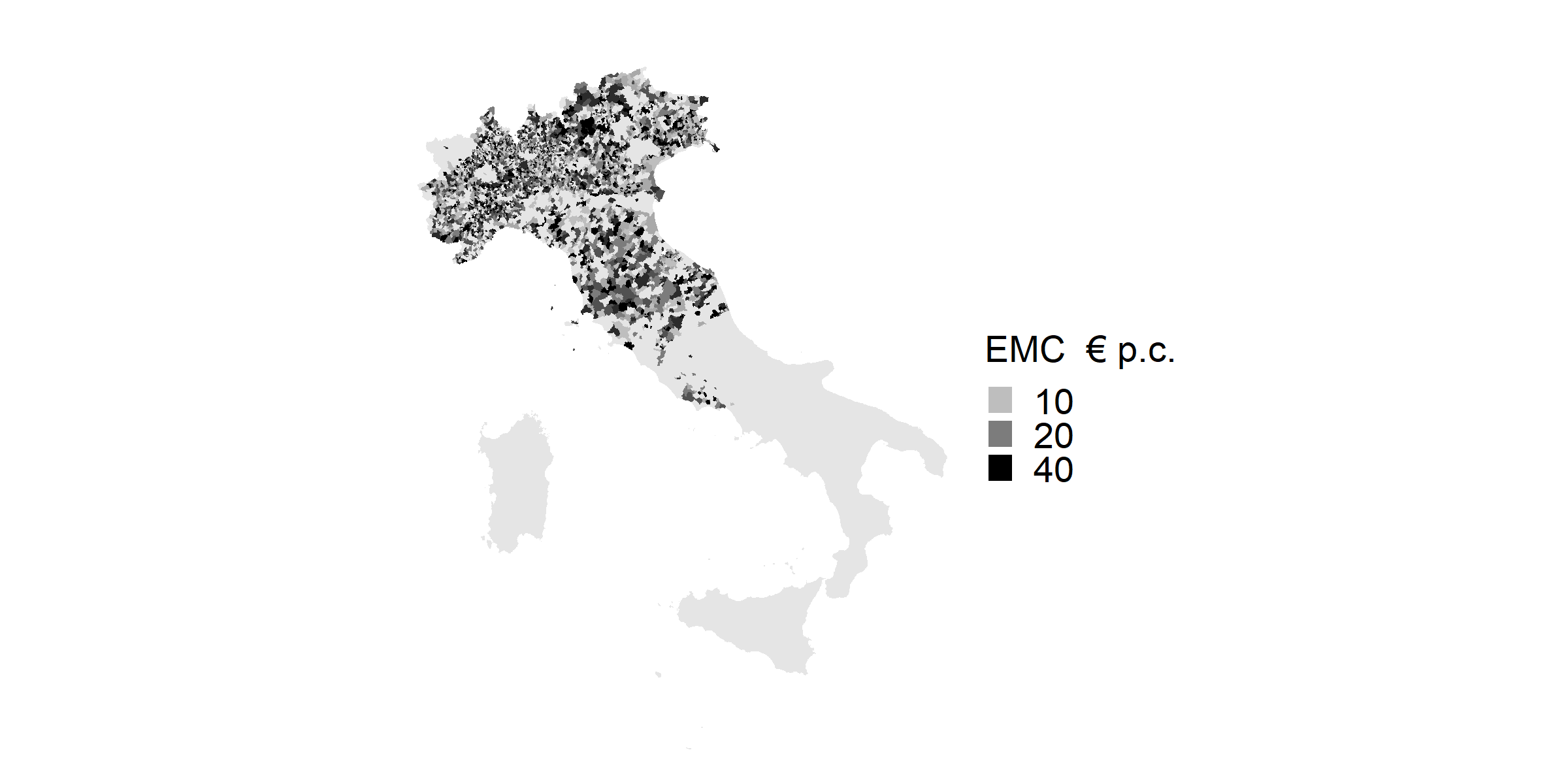}}
       \caption{ \footnotesize Distribution of predicted Effects on Municipal Costs (EMC) in the third policy year in \EUR{} per capita.}
 \footnotesize
 \label{fig:maplast}
    \end{figure}
         \end{center}

\begin{table}[H]
\centering
\footnotesize
\begin{tabular}{llrrr}
% \hline
 & Year  & Mean & Sd & EMC$>$0 \\ 
 \hline

 EMC \EUR{} p.c. &   1 & 1.45 & 10.90 & 50\%\\ 

& 2   & 3.95 & 9.33 & 73\% \\ 

 &3    & 23.85 & 18.33 & 99.8\% \\ 

   \hline
\end{tabular}
 \caption{ \footnotesize Summary statistics for predicted policy Effects on Municipal Costs (EMC) for all observed municipalities. EMC$>0$ indicate cost savings per capita (p.c.) by policy year.}
 \label{tab:welfarepred}
\end{table}
\noindent
Compared to a counterfactual zero-price (no PAYT) scenario, UW reductions lead to a 24\% decrease in municipal costs of waste management on average over the policy years. At the same time, increased RW raises municipal costs by on average 21\%. In year 3, average per capita total cost savings amount to \EUR{}24, however, there is large variation. Figure \ref{fig:maplast} maps this variation: despite effect heterogeneity, waste prices cause cost savings for most municipalities.

\subsection{Robustness and sensitivity}\label{sec:robustness}

I perform additional analyses to check whether my results are robust against the presence of unaccounted-for heterogeneity. In line previous findings \citep[see][for a review]{Bel2014}, I find that my estimates are robust to confounding from adoption of heterogeneous PAYT systems (weight versus volume). Moreover, I find that waste tourism or, more generally, spillover effects are not relevant on any significant scale. See Appendix  \ref{sec:robustness1} for more details.

Finally, I contrast my average estimates to the binary treatment case, two-way (event-study-like) fixed effects estimation (with/without re-weighting á la \citealt{Chaise2020}), and R-learning LASSO regression. Results highlight the importance to account for continuous rather than binary treatment, the bias of the fixed effects estimator due to non-parallel trends pre-policy, and robustness to the specific choice of R-learning estimator. Full results are discussed in Appendix \ref{sec:robustness2}. 
%%%%%%%%%%%%%%%%%%%%% CONCLUSIONS
\section{Conclusions} \label{Sec:conclusions}

This paper models and estimates causal effects of waste prices (PAYT) on recycling, unsorted, and total waste amounts by explicitly accounting for heterogeneity with machine learning methods. I analyze unique data with large variation in prices and a high-dimensional set of municipal characteristics. I estimate municipal level price elasticities using a random forest and R-learning estimator robust to confounding that affects outcomes and prices. By mapping municipal attributes onto causal effects, I predict heterogeneous policy effects for all municipalities in the sample, treated or not. The analysis of policy effects on municipal costs is motivated by a theoretical model that predicts heterogeneous waste behaviors in response to PAYT. Such heterogeneity in waste demands, coupled with the large
heterogeneity in municipal waste management costs, make the sign and magnitude of PAYT effects on municipal costs ex-ante ambiguous and, thus, an empirical question.

\setlength\parindent{0.5cm}
Disentangling effect heterogeneity in prices from other sources, I find that waste demands are nonlinear. Low prices make waste demands similarly elastic, and trigger large unsorted waste reductions, increase recycling, and decrease total waste. High prices make waste demands increasingly elastic, and induce recycling rather than total waste reductions. On average, policy causal effects amount up to -50\% on unsorted waste, 32\% on recycling, and -5\% on total waste. Income effects and waste habits pre-policy significantly explain price effect heterogeneity at high prices. I find that lower-income municipalities are more price elastic. Despite large effect heterogeneity, the policy reduces municipal costs in all municipalities after three years of adoption, especially where total waste reductions are large. This implies that higher prices may be less desirable from a municipal cost perspective as they do not cause further total waste reductions, instead they rather reallocate waste to the recycling pile. 
\setlength\parindent{0.5cm}
This study shows that price levels matter for waste behavior and municipal costs of waste management. Estimating heterogeneous responses on unsorted waste, and especially on recycling and total waste is crucial to assess overall policy effects on municipal costs. Reductions of total waste may not be immediate, and municipal costs may increase in the short-run. One upshot of this paper is that low prices can have a significant impact on waste behavior and municipal costs. Analyzing the long-term effects of PAYT policies presents an exciting research opportunity that I hope to tackle in future work.

%\section*{Acknowledgements} \label{Sec:ackn}

  \newpage
  \footnotesize
  \bibliography{main}
\bibliographystyle{chicago}

\newpage
\appendix

\section*{Online Appendix}

  \normalsize

\section{Proofs}

\subsection{The theoretical model} \label{Sec:appendix0}
Consider the Lagrangian (\ref{eq2lag}). The first-order conditions for the choice variables $UW, RW$ are: 
\begin{gather}    \label{foc1}  
 \alpha U_c/\lambda= p_c\alpha+p_{UW}\\   \label{foc2}
  \alpha U_c/\lambda= p_c\alpha+w\chi RW\\ \label{foc3}
 wH-w(\frac{\chi RW^2}{2})-p_{c}\alpha(RW+UW)-p_{UW}UW=0.
\end{gather}
\noindent
Solving (\ref{foc1}) and (\ref{foc2}) results in $p_{UW}=w\chi RW$ which says that households choose the optimal amount of recycling when the marginal cost of recycling equals the marginal cost of unsorted waste generation. The utility-maximizing solution for the choice of recycling is: 
\begin{equation} \label{RWopt}
    RW^*=\frac{p_{UW}}{w\chi}.
\end{equation}
Differentiating this equation with respect to $p_{UW}$ leads to the comparative static (\ref{cs2}): $\frac{\delta RW^*}{\delta p_{UW}}=\frac{1}{w\chi} > 0$.

The utility-maximizing solution for the choice of unsorted waste is obtained by substituting (\ref{RWopt}) in (\ref{foc3}): 
\begin{equation}
    wH-\frac{p_{UW}^2}{2w\chi}-p_{c}\alpha(\frac{p_{UW}}{w\chi}+UW^*)-p_{UW}UW^*=0
\end{equation}
Rearranging this equation in terms of $UW^*$ leads to:
\begin{equation} \label{eqlast}
UW^*(p_{c}\alpha+p_{UW})= wH -p_{c}\alpha \frac{p_{UW}}{w\chi}-\frac{p_{UW}^2}{2w\chi}.
\end{equation}
Differentiating (\ref{eqlast}) with respect to $p_{UW}$ (using the quotient rule) gives: 
\begin{equation} \label{eqlastlast}
\frac{\delta UW^*}{\delta p_{UW}} = \frac{-\frac{(p_{c}\alpha+p_{UW})}{w\chi}(p_{c}\alpha+p_{UW})-UW^*(p_{c}\alpha+p_{UW})}{(p_{c}\alpha+p_{UW})^2}.
\end{equation}
After simplifying this last expression and using (\ref{foc1}) to substitute for $(p_c\alpha+p_{UW})=\alpha U_c/\lambda^*$, we obtain the comparative static (\ref{cs1}): $\frac{\delta UW^*}{\delta p_{UW}}=-\frac{1}{w\chi}-\frac{UW^*}{\alpha U_c/\lambda} < 0$.

\subsection{The econometric model} \label{Sec:appendix00}
Consider the partially linear waste outcome model (\ref{eqY}). Let us take the conditional expectation with respect to $X_i$ on both sides, change signs, and add $Y_i$ to both sides. We obtain:
\begin{equation}\label{eqYresproof}
\underbrace{Y_i}_{added} -\e{Y_i|X_i}  = \underbrace{Y_i}_{added}-\e{Y_i(0)|X_i} - \e{P_i|X_i}\delta(X_i) - \e{\epsilon_i(P_i)|X_i},
\end{equation}
where $\e{\epsilon_i(P_i)|X_i}=0$. Substituting $Y_i$ on the right-hand side for the corresponding expression (\ref{eqY}) gives:
\begin{equation}\label{eqYresproof2}
Y_i -\e{Y_i|X_i}  =  \underbrace{{\e{Y_i(0)|X_i}} + P_i\delta(X_i) + \epsilon_i(P_i)}_{Y_i} {-\e{Y_i(0)|X_i}} - \e{P_i|X_i}\delta(X_i).
\end{equation}
After simplifying and rearranging this last expression we get equation (\ref{eqYres}), namely, $Y_i - \e{Y_i | X_i} = (P_i - \e{P_i | X_i})\delta(X_i) + \epsilon_i(P_i)$.

\section{Descriptive statistics} \label{Sec:appendixA}
\vspace{-1cm}
\renewcommand{\arraystretch}{1}
 \footnotesize
\begin{longtable}{ll}  \centering 

 \endfirsthead
 \endhead
    \hline \multicolumn{2}{r}{{Cont'd}} 
\endfoot
\endlastfoot
  \caption{Variables' description. Census indicates 2011 values \citep{ISTATcensus}.} \\
  \label{table:var_desc} \\[-2.8ex]
  \hline
 & Variables' description \\ 
  \hline
    PAYT price & Treatment: unit price on unsorted waste in \EUR{} per liter\\
Recycling Waste (RW) kg p.c. & Recycling waste per capita (kg) \\ 
  Unsorted Waste (UW) kg p.c.  & Unsorted waste per capita (kg) \\ 
  Total Waste (TW=UW+RW) & Total waste per capita (kg) \\ 
  UW management costs p.c & Per capita costs of UW management (\EUR{}) \\ 
  RW management costs p.c & Per capita costs of RW management less RW revenues (\EUR{}) \\ 
  UW unit costs per kg & Unit costs of UW management (\EUR{} per kg) \\ 
  RW unit costs per kg & Unit costs of RW management less RW revenues (\EUR{} per kg) \\ 
  Km to PAYT city & Distance to closest municipality with PAYT in t-1 (km) \\ 
  Km to hazardous waste site & Distance to closest hazardous waste treatment facility (km) \\ 
  Km to incinerator & Distance to closest waste incinerator (km) \\ 
  Km to landfill & Distance to closest waste landfill (km) \\ 
  Population density & Population density (inhabitants per km$^2$) \\ 
  Household size & Average household size (n. household members) \\ 
  Income p.c. & Income per capita (x \EUR{}1,000) \\ 
  Net migration p.c. & Net migrant flow per capita \\ 
  Population & Population (x 1,000 inhabitants) \\ 
  Share of foreign population & Share of foreign population \\ 
  Share of male population & Share of male population \\ 
  Population growth & Population growth \\ 
  Tourism & Capacity of tourist accommodation per capita (x 1,000) \\ 
  Share of population aged $<5$ & Share of population aged less than 5 years old (census) \\ 
  Share of population aged $<14$ & Share of population aged less than 14 years old (census) \\ 
  Share of population aged $>65$ & Share of population aged more than 65 years old (census) \\ 
  Pop. share w. elementary deg. & Share of pop. with elementary degree or lower (census) \\ 
  Pop. share w. college deg. & Share of population with college degree (census) \\ 
  Share of rented houses & Share of rented houses (census) \\ 
  Housing density & Housing density (inhabitants per 100m$^2$, census) \\ 
  Share of single-parent families & Share of single-parent families (census) \\ 
  Share of students aged $>15$ & Share of student population older than 15 years old (census) \\ 
  Share of commuters & Share of commuters (census) \\ 
  Social deprivation index & Social deprivation index (-/+ deprived, census) \\ 
  Out of labor force/active pop. & Out of labor force pop./pop. aged 15-64 years old (census) \\ 
  Unemployment rate & Unemployment rate (census) \\ 
  Commuting intensity & Commuting index (0-100, IIRFL) between cities (census) \\ 
  Political participation & Voter turnout in the 2013 Italian general election (IGR13) \\ 
  Vote share big-tent party & Vote shares for big-tent parties in the IGR13 \\ 
  Vote share extreme left & Vote shares for extreme left-wing parties in the IGR13 \\ 
  Vote share extreme right & Vote shares for extreme right-wing parties in the IGR13 \\ 
  Local mayor (0/1) & Mayor born in the municipal province (dummy) \\ 
  Centre-party mayor (0/1) & Centre-party mayor (dummy) \\ 
  Green-party mayor (0/1) & Green-party mayor (dummy) \\ 
  Left-party mayor (0/1) & Left-wing mayor (dummy) \\ 
  Other-party mayor (0/1) & Mayor of other party (dummy) \\ 
  Local-party mayor (0/1) & Mayor of local party (dummy) \\ 
  Right-party mayor (0/1) & Right-wing mayor (dummy) \\ 
  Mayor's age & Mayor's age \\ 
  Mayor's years of office & Mayor's term of office (years) \\ 
  Province or region capital (0/1) & Either province or region capital (dummy) \\ 
  Province capital (0/1) & Province capital (dummy) \\ 
  Region capital (0/1) & Region capital (dummy) \\ 
  Medium urbanization (0/1) & Mediumly urbanized municipality (dummy) \\ 
  High urbanization (0/1) & Highly urbanized municipality (dummy) \\ 
  Low urbanization (0/1) & Lowly urbanized municipality (dummy) \\  
   \hline
   
\end{longtable}

  \footnotesize

\section{The overlap assumption} \label{Sec:appendixB}

  \normalsize
  
 \begin{figure}[H]
        \centering
\subfloat {\includegraphics[width=16cm,height=5cm,keepaspectratio]{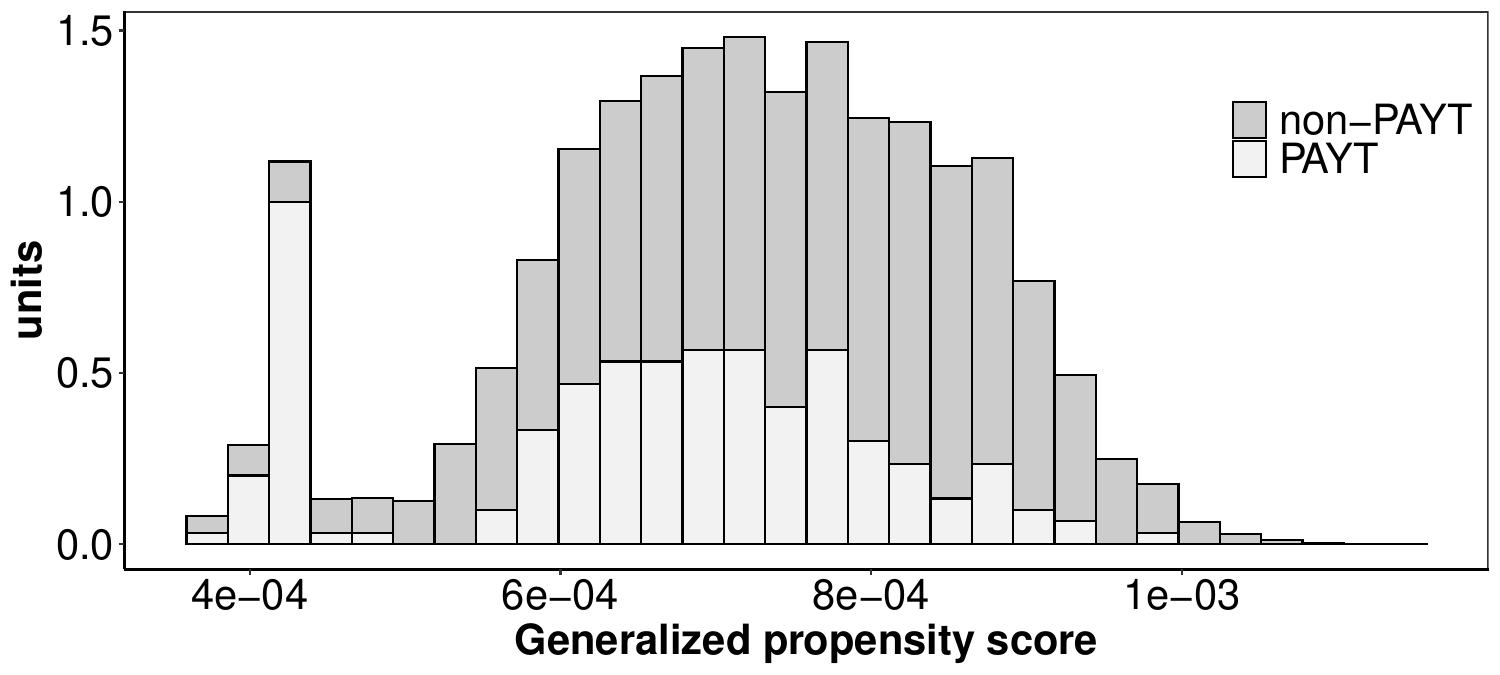}}
       \caption{ \footnotesize Common support condition for treated (PAYT) and never-treated (non-PAYT).}
  \label{fig:overlap}  \footnotesize
    \end{figure}

    \normalsize

  \section{Additional results} \label{Sec:appendixC}

\renewcommand{\arraystretch}{1.2}

    \subsection{Testing for effect heterogeneity}  \label{Sec:appendixC:het}

Levene's tests (\citeyear{Levene1960}) reject the null hypothesis of no CAPE heterogeneity for all outcomes. Moreover, as suggested in \citet{athey2017}, I perform the following heuristic to test for effect heterogeneity. First, I group observations into a high and low APE subgroup using the median CAPE as a threshold. Next, I derive an estimate of the APE for each subgroup by residual-on-residual regression. Table \ref{tab:ATEdiff} shows statistically significant differences in APE between subgroups for all outcomes and policy years.

\begin{table}[H]
\centering \footnotesize
\begin{tabular}{llllllllll}
  \hline
 Year & $\lvert dUW \rvert$ & ci.low & ci.up & $\lvert dRW \rvert$ & ci.low & ci.up & $\lvert dTW \rvert$ & ci.low & ci.up \\ 
   \hline
   1 & 2.72 & 2.46 & 2.98 & 5.05 & 4.82 & 5.29 & 5.77 & 5.51 & 6.04 \\ 
  2 & 3.69 & 3.43 & 3.95 & 5.15 & 4.91 & 5.38 & 12.01 & 11.62 & 12.41 \\ 
  3 & 4.49 & 4.24 & 4.75 & 6.13 & 5.91 & 6.35 & 3.32 & 3.07 & 3.57 \\ 
  \hline
\end{tabular}
 \caption{ \footnotesize \centering Absolute difference in APE ($\lvert dUW \rvert$, $\lvert dRW \rvert$, $\lvert dTW \rvert$ in kg) between high and low APE subgroup with 95\% confidence intervals [ci.low;ci.up].}
  \label{tab:ATEdiff}  \footnotesize
\end{table}

    \subsection{CAPE heterogeneity across price levels} \label{app:capekg}

\begin{figure}[H]
   \centering
\subfloat{\includegraphics[scale=0.3]{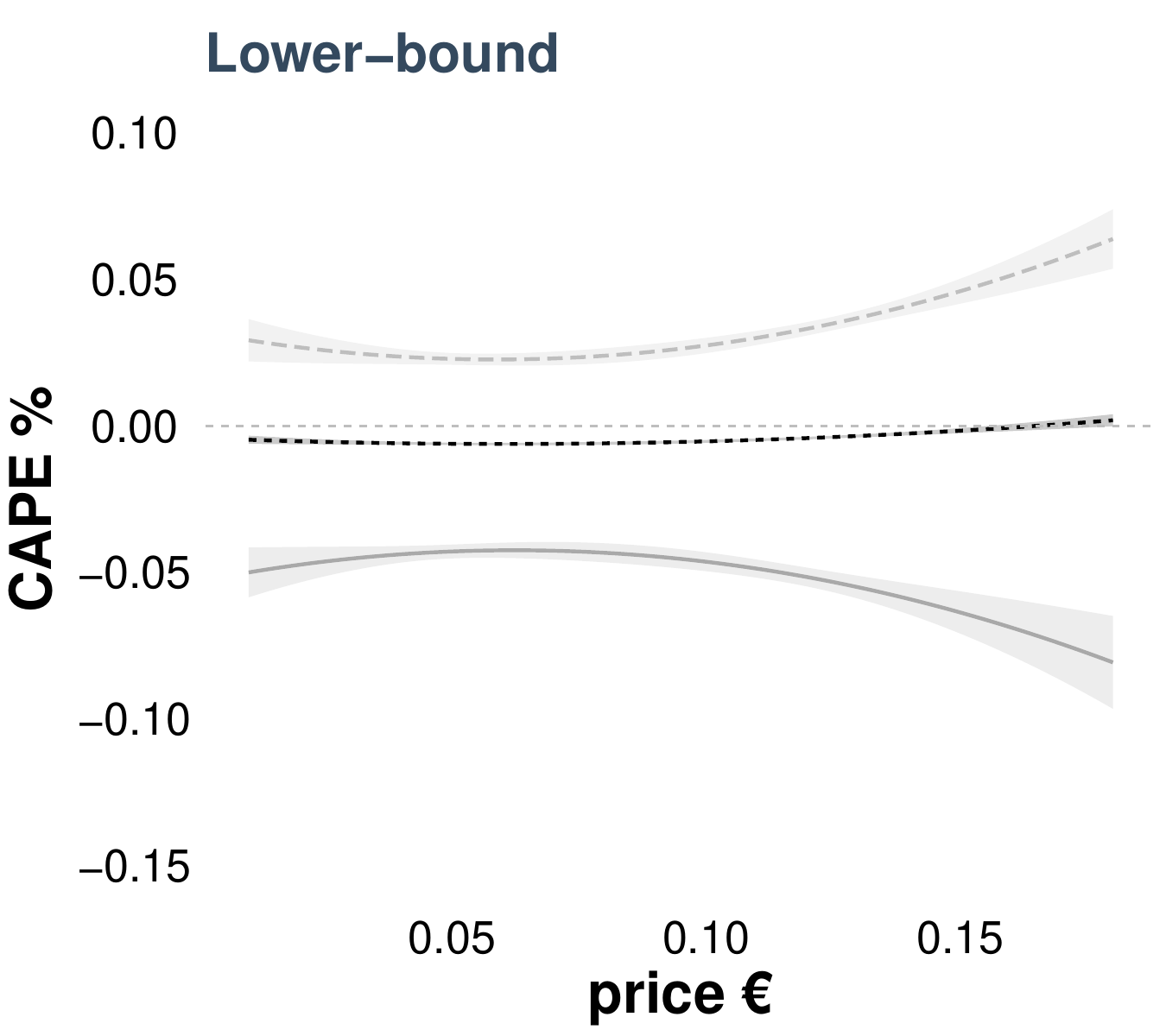}}

\subfloat{\includegraphics[scale=0.3]{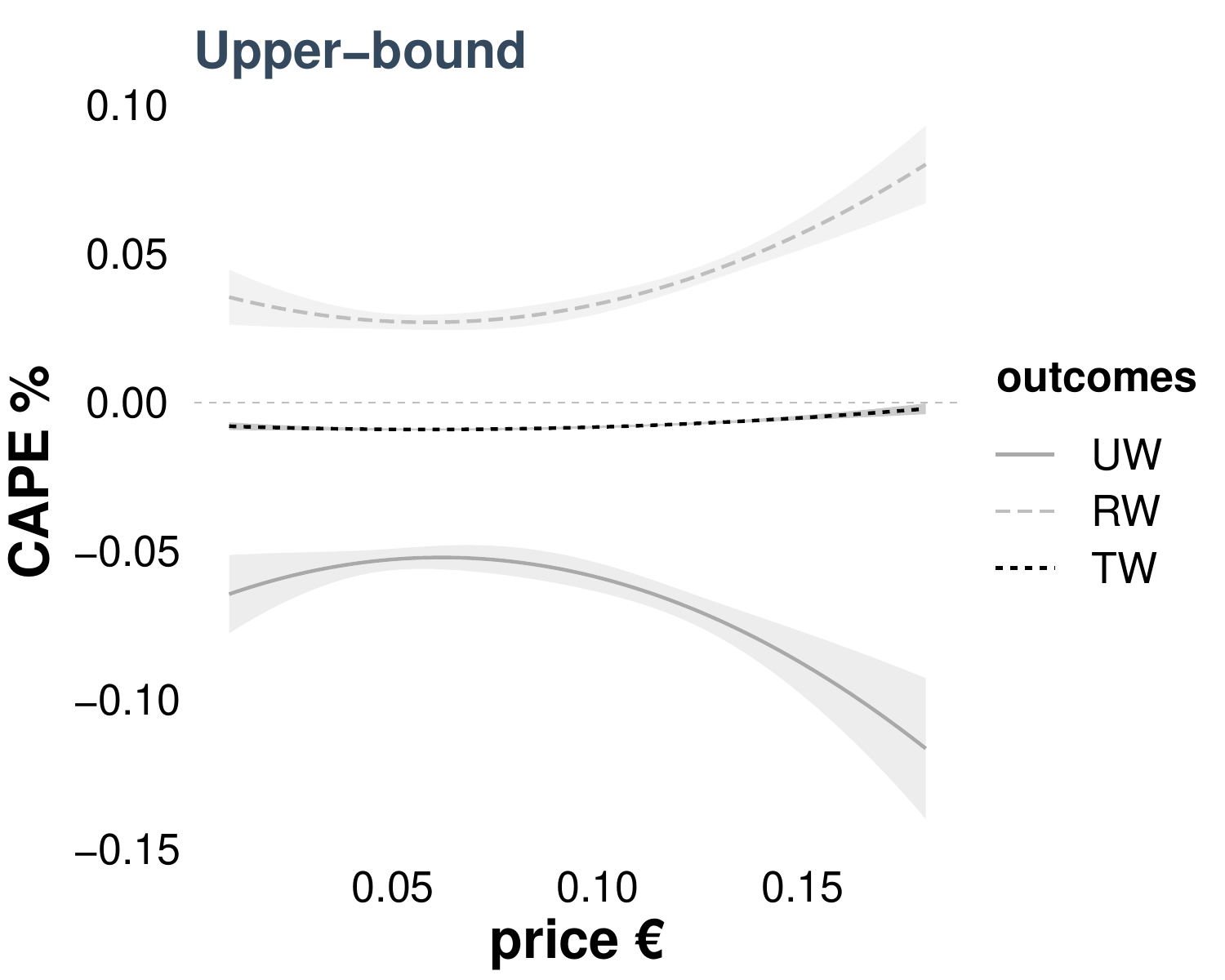}}
       \caption{ \footnotesize Fitted CAPE of waste demands at each price level. Fitted CAPE in the left (right) panel are estimated using CAPE lower-bounds (upper-bounds) of 95\% confidence intervals.}
 \label{fig:uncertaintyCAPE}
    \end{figure}

Figure  \ref{fig:uncertaintyCAPE} reports fitted lower- and upper-bound CAPE estimates at each price level. Lower- and upper-bounds refer to the 95\% confidence intervals estimated in the RF procedure via bootstrap of little bags \citep{atheyGRF, laake2009}.  Figure  \ref{fig:uncertaintyCAPE} shows that the shape and magnitude of the heterogeneity do not largely vary when accounting for uncertainty bounds of the CAPE point estimates.

     \normalsize
 \noindent
Figure \ref{fig:mainkg} reports effect heterogeneity across price levels, ceteris paribus. Price effects on the y-axis (CAPE kg) are estimated as quantity changes for a one cent price increase. CAPE heterogeneity varies at low versus high prices, ceteris paribus: a one cent price increase reduces UW by on average 10.8 kg at low prices vs. 14.2 kg at high prices, ceteris paribus. Higher prices lead to increased recycling by on average 6.8 kg at low prices vs. 11.6 kg at high prices. The remaining UW reductions are due to TW reductions.

\begin{center}

    \begin{minipage}[H]{\linewidth}
      \begin{figure}[H]
   \centering
\subfloat{\includegraphics[scale=0.35]{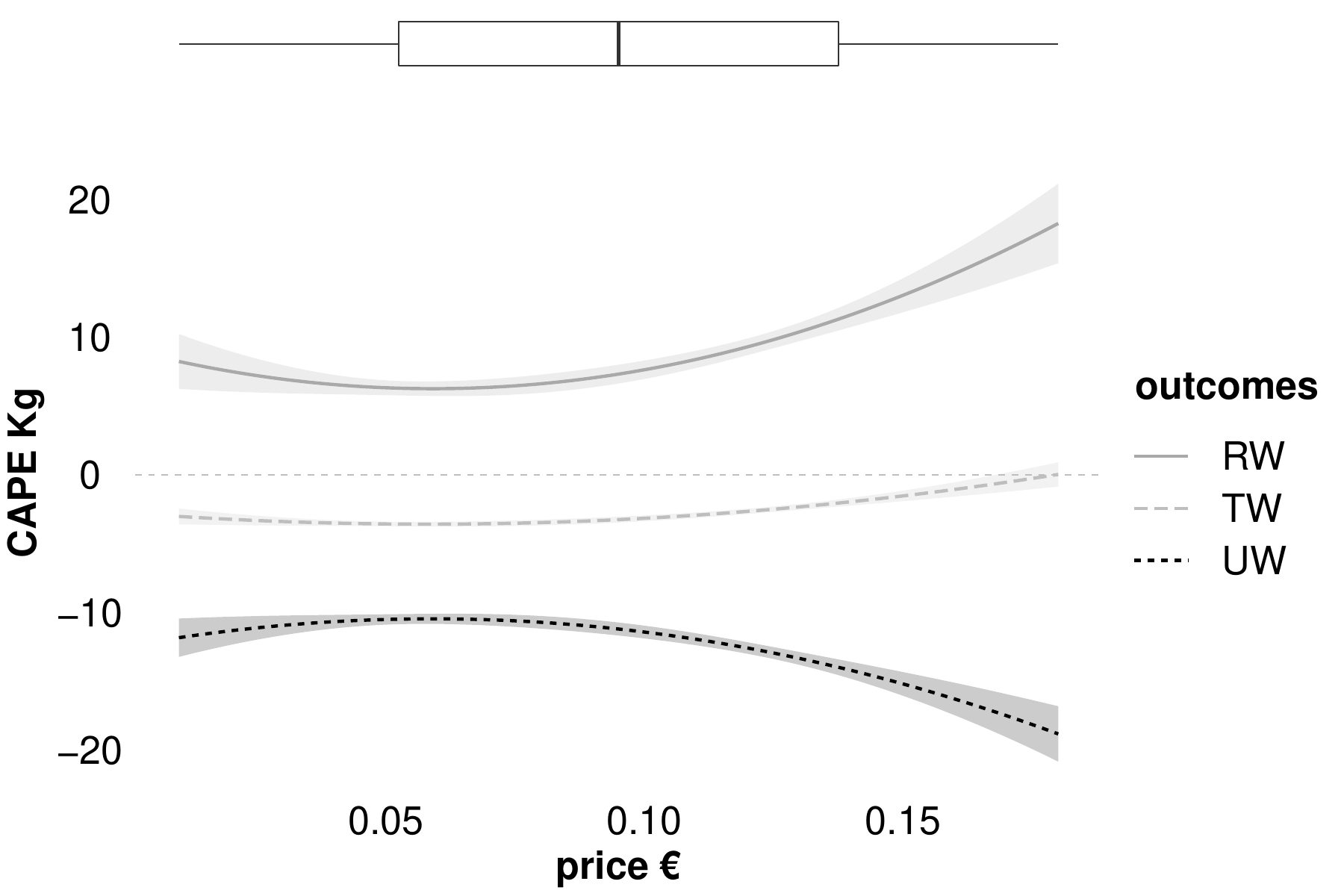}}
       \caption{ \footnotesize Fitted price effects on waste demands (CAPE kg) as quantity changes for a one euro cent price increase.}
 \footnotesize
 \label{fig:mainkg}
    \end{figure}
  \end{minipage}
  \end{center}
       \normalsize
 \noindent
Figure \ref{fig:mainRW} reports effect heterogeneity across prices by pre-policy RW levels, ceteris paribus. Effect heterogeneity is statistically significant only at high prices: municipalities recycling little before policy have generally higher elasticities, ceteris paribus, by on average 1.4 percent points.

\begin{center}

    \begin{minipage}[t]{\linewidth}
      \begin{figure}[H]
   \centering
\subfloat{\includegraphics[scale=0.48]{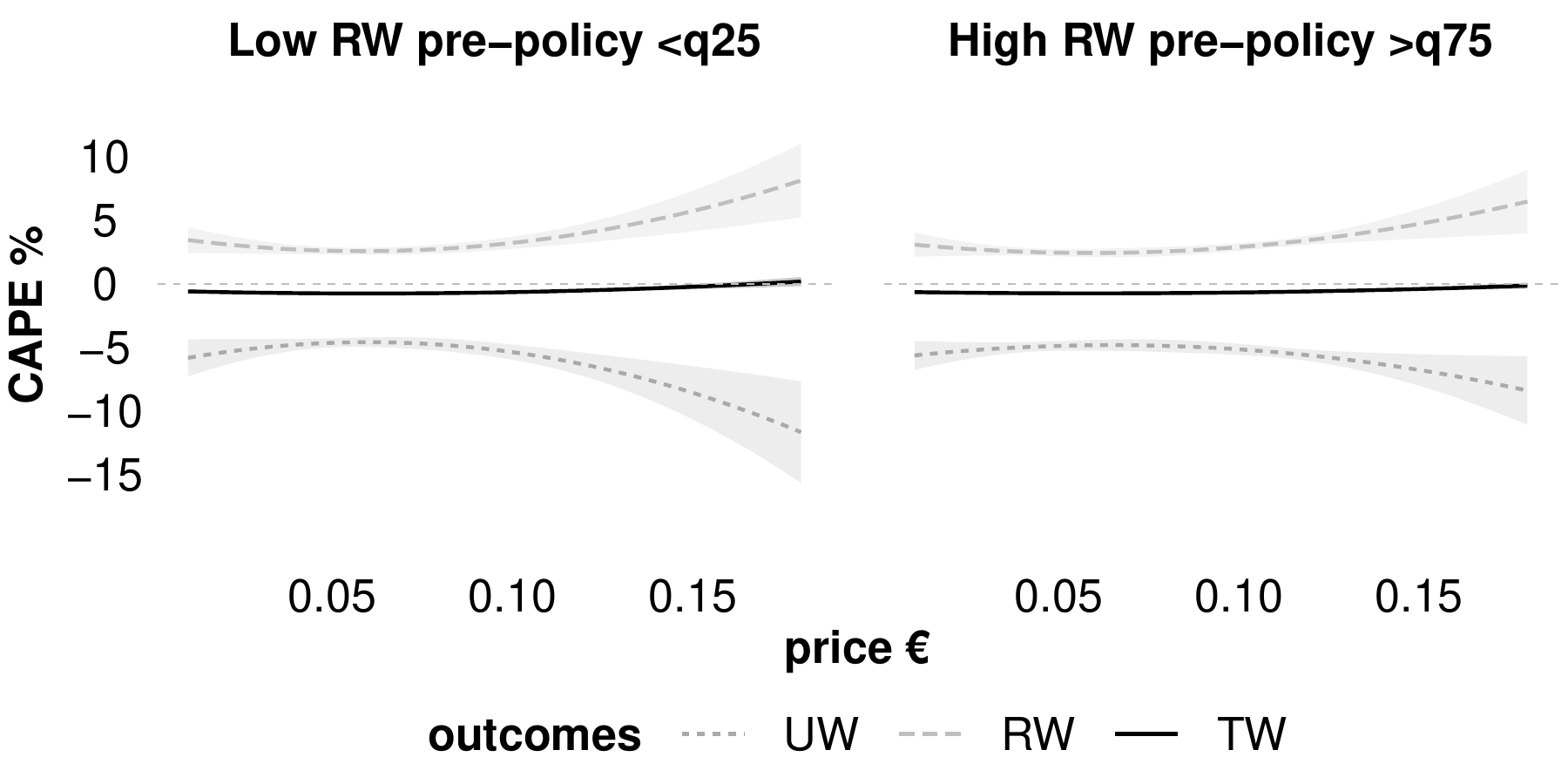}}
       \caption{ \footnotesize Fitted price semi-elasticities (CAPE) of waste demands by pre-policy recycling levels. Thresholds q25 and q75 indicate first (248 kg) and third (344 kg) quartiles of annual recycling per capita.}
 \footnotesize
 \label{fig:mainRW}
    \end{figure}
  \end{minipage}
  \end{center}
\noindent

      \begin{figure}[H]
   \centering
\subfloat{\includegraphics[scale=0.25]{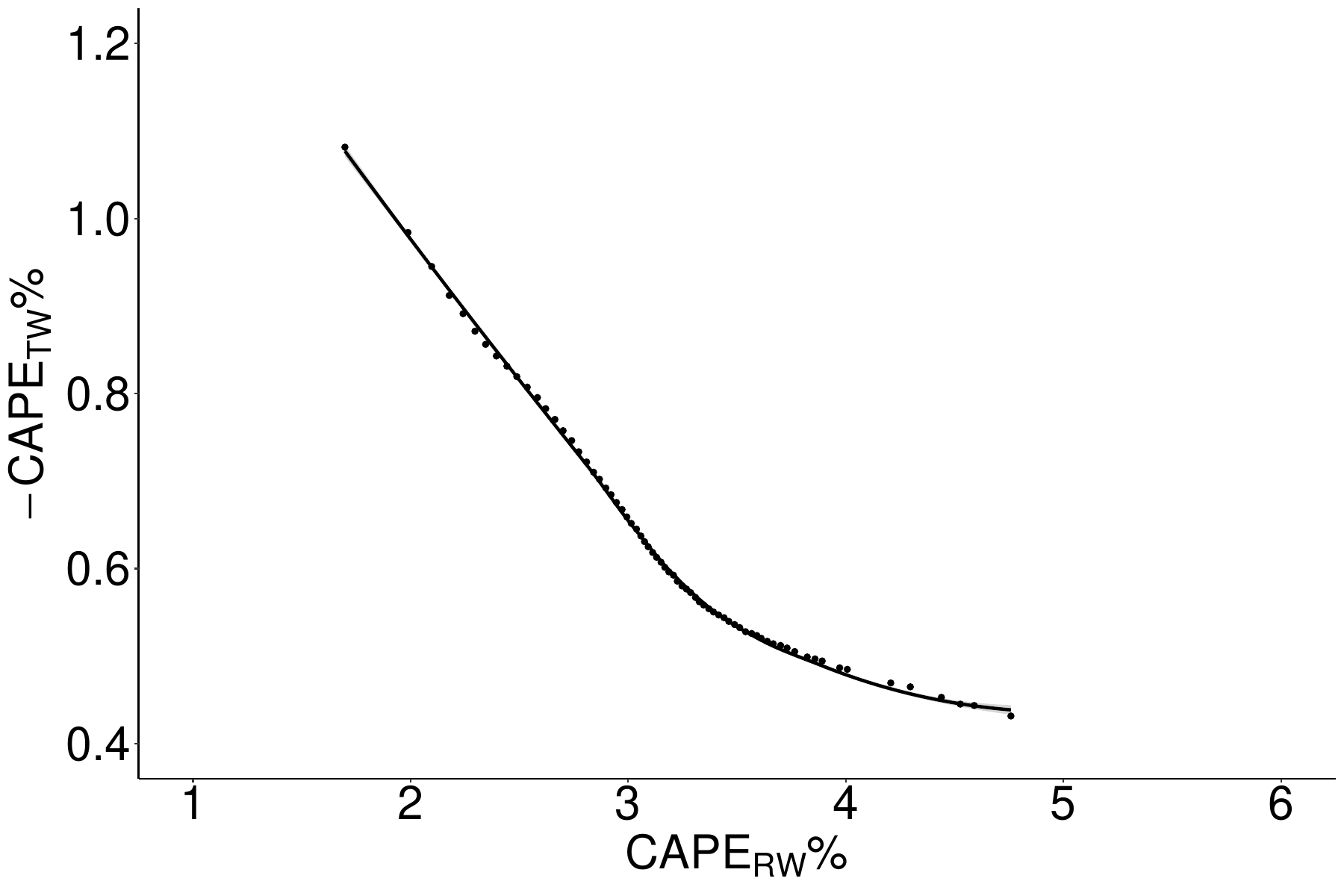}}
       \caption{ \footnotesize CAPE (semi-elasticities) for recycling ($x$-axis) plotted against CAPE for total waste reductions ($y$-axis). A smooth line is fitted through percentile values.}
 \footnotesize
 \label{fig:cape5}
    \end{figure}

 \subsection{Distribution of policy effects on municipal costs}\label{Sec:appendixD:swe}

    \begin{minipage}[t]{\linewidth}
      \begin{figure}[H]
   \centering
\subfloat{\includegraphics[scale=0.4]{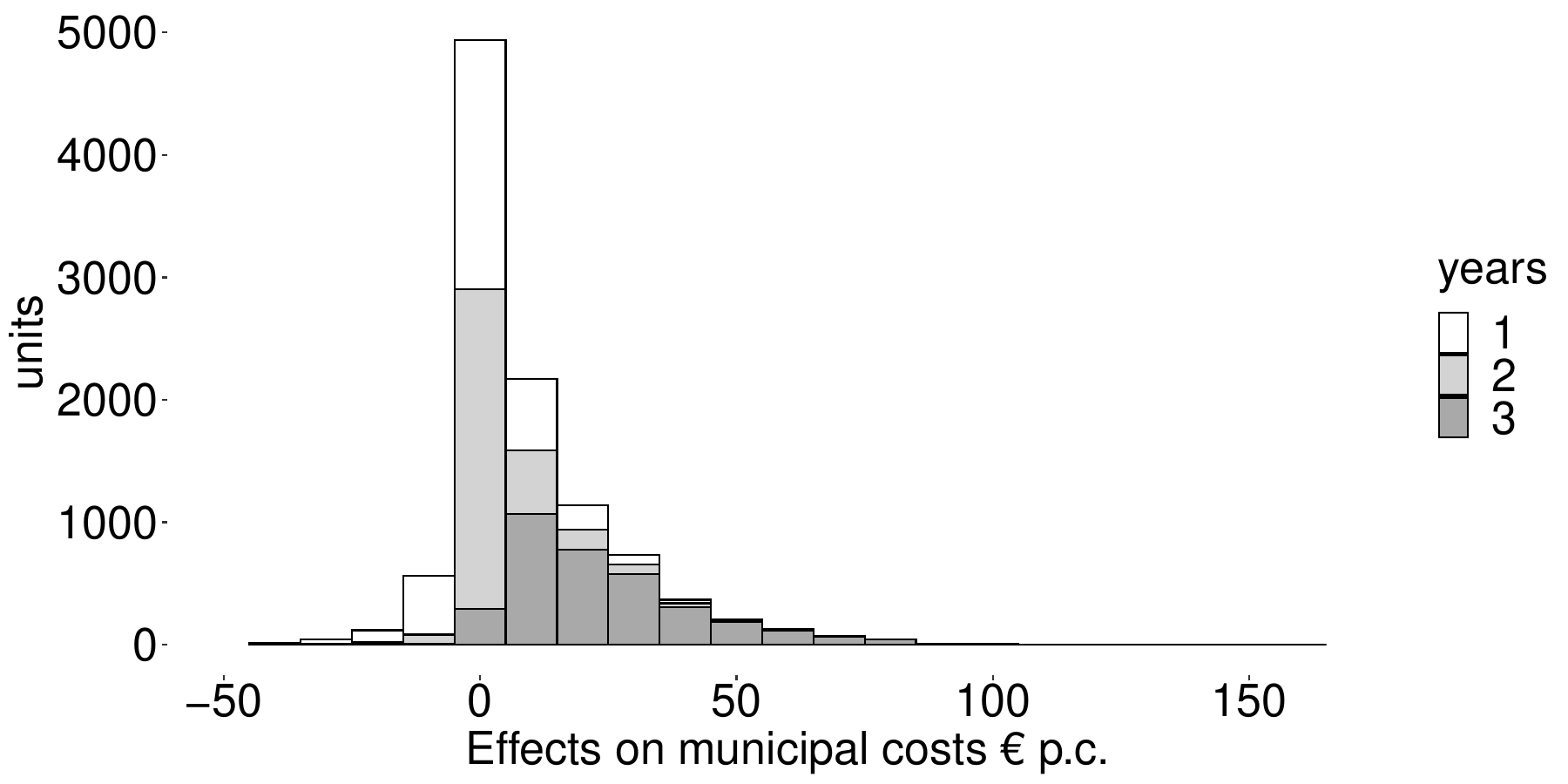}}
       \caption{\footnotesize Distribution of policy Effects on Municipal Costs (EMC$>$0 indicate cost savings) for all observed municipalities after one, two, and three years of policy adoption  in \EUR{} per capita. The y-axis refers to the number of treated and yet-to-be-treated municipalities.}
 \footnotesize
 \label{fig:hist_swe}
    \end{figure}
  \end{minipage}

 \section{Robustness and Sensitivity}

\subsection{Possible unaccounted-for heterogeneity and spillovers} \label{sec:robustness1}

  \normalsize
\textit{Waste tourism.}--Violation of the no spillover assumption (SUTVA) could possibly come from waste tourism, namely, households may discharge their waste for free in surrounding municipalities without PAYT. I geocoded neighbors as never-treated units sharing a border with treated units. The remaining untreated units represent the control group. In order to estimate causal effects of PAYT on waste amounts of surrounding municipalities without PAYT, I use the RF running the R-learner described in Section \ref{Sec:method}. I find that PAYT has no significant effect on the waste amounts collected in surrounding municipalities on average. As a further robustness check, I re-estimate causal effects for PAYT municipalities excluding neighbors from the control group, finding no significant change in estimates. As in other empirical studies \citep[e.g.][]{carattini2018, Allers2010,Dijkgraaf2004}, waste tourism is not relevant on any significant scale.

\vspace{0.5cm}
\textit{System effects.}-- Unaccounted-for heterogeneity may threaten identification. Adoption of different PAYT systems (e.g. volume versus weight) may have heterogeneous effects on waste generation.\footnote{I cannot include system dummies in RF because, as for any policy indicator, I would force treated and untreated units into different neighborhoods.} In the sample, about 7\% of treated units pay per weight, 16\% per bag, 72\% per emptying, and 5\% have a mixed system. I first test for whether the included covariates can explain system adoption. Second, I test for significant differences in price elasticities between systems. 

Using all covariates, I predict system choices by RF for classification. I find that the included observables can predict (out of sample) system choices correctly for most municipalities.  How do municipal characteristics affect system choices? Evidence provided by manufacturers of PAYT technology (e.g., magnetic cards, tags and RFID readers) suggests that system choices are driven by three main factors. First, proximity: neighboring municipalities may either share the same waste collection company or influence each other through information dissemination. Second, population density: sparsely populated municipalities are more likely to adopt weight systems \citep{partitalia}.\footnote{Implementing curbside collection is especially difficult in low-density areas. Also, transportation costs would be high. Thus, communities generally prefer to organize waste collection in centralized disposal areas rather than door-to-door. As volume systems would imply carrying heavy bags and bins, weight systems facilitate households allowing them to carry smaller quantities of waste.} Third, geography: physical characteristics of the territory such as uphills and one-way roads impact transportation costs and collection modes.

\setlength\parindent{0.5cm}
I run a multinomial logit regression in order to assess whether observables can partly explain the choice of the adopted PAYT system. These are: population size (Population), distance to neighboring PAYT municipalities (Km to PAYT city), population density (Population density), and urbanization level (Medium/High/Low) as described in Table \ref{table:var_desc} in Appendix \ref{Sec:appendixA}. In particular, I specify urbanization levels as urban, peri-urban, semi-peripheral, and very peripheral (the reference category) based on travel times (t) from each municipality to the closest urban center. Table \ref{tab:multinom} shows the results including all treated observations in the first policy year. The dummy for weight systems is the reference choice.

\renewcommand{\arraystretch}{0.9}

  \footnotesize
\begin{table}[H] \centering 
  \footnotesize
  \caption{Average partial effect estimates of multinomial logit regression. Weight systems are the reference choice. Very peripheral areas (t$>$75') are the reference category for geographic dummies.} 
  \label{tab:multinom} 
   \footnotesize
\begin{tabular}{@{\extracolsep{5pt}}lccc}    
 \footnotesize
 & \multicolumn{3}{c}{\textit{Dependent variable:}} \\ 
\cline{2-4} 
\\[-1.8ex] & emptying & bag & mixed \\ 
\\[-1.8ex] & (1) & (2) & (3)\\ 
\hline \\[-1.2ex] 
 Population & $-$0.085$^{***}$ & $-$0.076$^{**}$ & 0.032 \\ 
  & (0.032) & (0.036) & (0.022) \\ 
  & & & \\ 

 Population density & 0.009$^{*}$ & 0.011$^{**}$ & 0.011$^{**}$ \\ 
  & (0.005) & (0.005) & (0.005) \\ 
  & & & \\ 
 Km to PAYT city & $-$0.006 & $-$0.017$^{*}$ & $-$0.073 \\ 
  & (0.008) & (0.010) & (0.046) \\ 
  & & & \\ 
   Peripheral(40\textless t\textless 75') & 1.189$^{*}$ & 1.655$^{***}$ & 28.377$^{***}$ \\ 
  & (0.663) & (0.000) & (0.663) \\ 
  & & & \\ 
Semi-peripheral(20\textless t\textless 40') & $-$21.713$^{***}$ & 5.599$^{***}$ & 4.797$^{***}$ \\ 
  & (0.924) & (0.952) & (0.962) \\ 
  & & & \\ 
Semi-urban(t\textless 20') & $-$24.078$^{***}$ & 2.523$^{***}$ & 2.132$^{***}$ \\ 
  & (0.699) & (0.718) & (0.781) \\ 
  & & & \\ 
Peri-urban & $-$24.819$^{***}$ & 1.111$^{**}$ & $-$36.060$^{***}$ \\ 
  & (1.618) & (0.525) & (0.000) \\ 
  & & & \\ 
   Urban & $-$20.275$^{***}$ & 4.625$^{***}$ & $-$5.571$^{***}$ \\ 
  & (1.097) & (0.912) & (0.223) \\ 
  & & & \\ 

 Constant & 36.339$^{***}$ & 13.313$^{***}$ & 0.799 \\ 
  & (1.145) & (0.980) & (0.706) \\ 
  & & & \\ 
\hline \\[-1.8ex] 
Observations & 194 & 194 & 194 \\
Year and area dummies & Yes & Yes & Yes \\
Akaike Inf. Crit. & 284.558 & 284.558 & 284.558 \\ 
\hline \\[-1.8ex] 
\textit{Note:}  & \multicolumn{3}{r}{$^{*}$p$<$0.1; $^{**}$p$<$0.05; $^{***}$p$<$0.01} \\ 
\end{tabular} 
\end{table} 

\noindent
\normalsize
Results show that lower population density (Population density) significantly increases the probability to adopt weight programs over any other, as expected. Higher populated municipalities (Population) are more likely to implement weight vs. bag/emptying programs, perhaps to exploit returns to scale. Information dissemination effects (Km to PAYT city) positively increase the choice of bag programs. Less peripheral municipalities are more likely to adopt bag systems over weight systems, and highly urbanized municipalities are more likely to adopt weight systems over emptying or mixed systems. \\

Next, I test for effect heterogeneity. Table \ref{tab:tukey} reports pairwise mean comparison tests \citep{Games1976}. 

\renewcommand{\arraystretch}{1.2}

\begin{table}[H]
\centering
\footnotesize
\begin{tabular}{lrrrrrrrrr}
 % \hline
   \footnotesize
 PAYT system & dUW & s.e. & p-value & dRW & s.e. & p-value & dTW & s.e. & p-value \\ 
  \hline
emptying - bag & 0.12 & 0.27 & 0.96 & 0.47 & 0.24 & 0.18 & 0.55 & 0.23 & 0.06 \\ 
  mixed - bag & 0.19 & 0.52 & 0.98 & 0.56 & 0.46 & 0.59 & 0.62 & 0.44 & 0.47 \\ 
  weight - bag & 0.76 & 0.52 & 0.44 & 0.16 & 0.46 & 0.98 & 1.06 & 0.44 & 0.07 \\ 
  mixed - emptying & 0.07 & 0.48 & 1.00 & 0.09 & 0.42 & 1.00 & 0.07 & 0.40 & 1.00 \\ 
  weight - emptying & 0.64 & 0.48 & 0.52 & -0.31 & 0.42 & 0.87 & 0.50 & 0.40 & 0.57 \\ 
  weight - mixed & 0.57 & 0.66 & 0.81 & -0.40 & 0.57 & 0.89 & 0.44 & 0.55 & 0.85 \\ 
   \hline
\end{tabular}
\caption{\footnotesize Pairwise comparison of price elasticities (CAPE) on waste amounts by PAYT system: Mean differences (dUW, dRW, dTW in kg per euro cent), standard errors and p-values estimated by Tukey-Kramer's method (\citeyear{Kramer1956}).}\label{tab:tukey}
\end{table}
  \normalsize
\noindent
Mean differences are overall statistically insignificant at conventional confidence levels. Estimates only suggest that bag-based systems are (weakly) associated to larger reductions of total waste (significant at the 10\% level), namely, price semi-elasticities on total waste are 0.55 and 1.06 kg higher (in absolute value) than with emptying- and weight-based systems, respectively. Overall, my study does not point towards significant differences in price elasticities between PAYT systems.

\subsection{Robustness to alternative assumptions} \label{sec:robustness2}

I compare my average estimates to \textit{(i)} the binary treatment case, which assumes no effect heterogeneity in prices; \textit{(ii)} the two-way (event-study-like) fixed effects regression, which assumes no effect heterogeneity, and selection based on time-constant variables; and \textit{(iii)} penalized linear regression methods known as LASSO, which assumes no effect heterogeneity, sparsity and linearity in a high-dimensional set of covariates. 
\vspace{0.5cm}
\textit{Binary treatment case.}--The binary case assumes that the policy has homogeneous effects across prices, and that policy adoption is homogeneous in prices. Table \ref{tab:ATEbin} reports average price effects (APE) estimated as average effects of a binary treatment at the mean price level ($ATE/\bar{P}$). 
 
 \renewcommand{\arraystretch}{1.2}

\begin{table}[H]
\centering \footnotesize
\begin{tabular}{lll|llllll}
  %\hline
  &   \multicolumn{1}{l}{\textit{Continuous}} & & \multicolumn{2}{l}{\textit{Binary}}  & \\
  \hline
  Year & UW  kg & s.e.  & UW kg & s.e. \\
  \hline
1 & -7.38 & 1.30   &-7.60& 2.55  \\ 
   2 & -8.86 & 1.03 & -11.3& 1.97   \\ 
   3  & -11.5 & 1.53 & -16.5 & 1.08   \\ 
  \hline
\end{tabular}
 \caption{\centering \footnotesize Continuous APE vs. Binary APE estimates (kg) at the mean price (ATE/$\bar{P}$) from (level-level) residual-on-residual regression with year clustered standard errors.}
\label{tab:ATEbin}  \footnotesize
\end{table}

Results show that mean elasticities (APE) on UW are overestimated by 28\% on average relative to the continuous treatment case. Why does this happen? The forest kernel assigns similarity weights to units based on a residual-on-residual regression that ignores the price variation and uses a binary indicator instead. Hence, the algorithm may place units adopting high and low prices in the same neighborhood, and assign the same prediction to both units. However, I have shown that price responses at high and low prices are not the same, and the sources of effect heterogeneity differ.

\begin{figure}[H]
       \centering
\subfloat {\includegraphics[width=15cm,height=4cm,keepaspectratio]{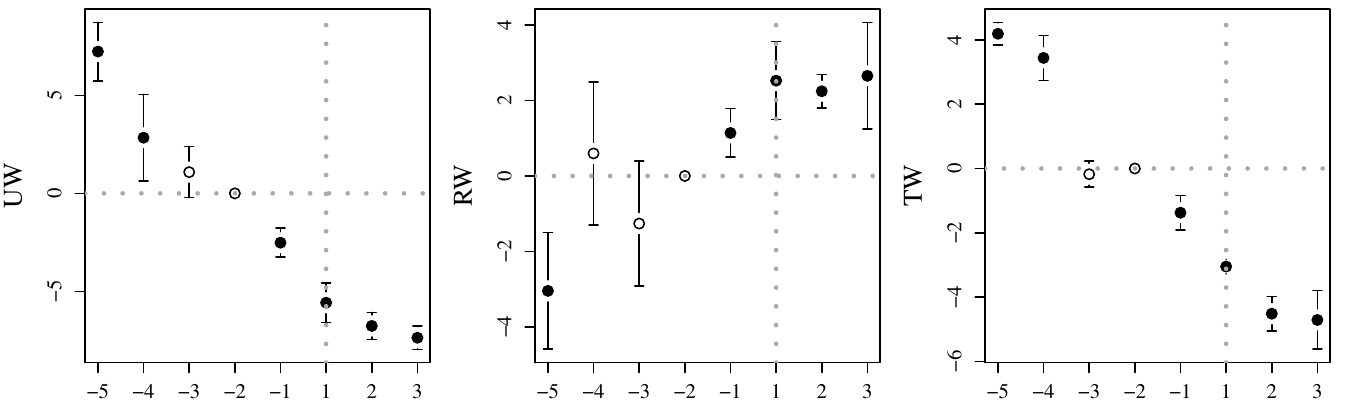}}
      \caption{ \footnotesize Dynamic DiD estimates (2010-2015, $n=19,982$) of average price effects (APE kg) on UW, RW, TW with year and unit fixed effects. Clustered standard errors \citep{Driscoll1998}. Statistical significance (5\%) indicated by black dots.}
  \label{tab:did}  \footnotesize
    \end{figure}

\vspace{0.5cm}
\textit{Two-way fixed effects regression.}--When the timing of treatment varies across units, Difference-in-Differences (DiD) designs are commonly extended to allow for dynamic average treatment effects by including leads and lags of treatment as regressors \citep{Jacobson1993}. These dummy variables allow policy effects to vary by the number of years relative to separation from policy. Figure \ref{tab:did} presents the results from a dynamic (event-study-like) DiD regression estimated using a standard two-way (unit and time) fixed effects model.\footnote{For the analysis, I use the software R, and, particularly, the package plm \citep{plm}.} The regression includes three leads and lags, and uses the second (unaffected) lead before policy (-2) as a baseline dummy. Since potentially relevant covariates are many and partly collinear, and there is no a priori guidance on which one to exclude, this DiD only controls for time-invariant waste generation determinants captured by municipal fixed effects. Figure \ref{tab:did} plots lead and lagged average price effect (APE) estimates (continuous treatment) with their confidence intervals. Black dots represent statistically significant effects at $5\%$, and vertical dotted lines indicate the policy adoption year.

\noindent
Results show that APE on UW are underestimated by 29\% on average (see first panel of Figure \ref{tab:did}). Moreover, standard event study pretrend tests \citep{Autor2003} present evidence of non-parallel trends, thus, a bias in the estimation of average price elasticities.\footnote{Including time trends in the regression does not induce parallel trends either.} There are at least two possible sources of bias. First, unaccounted-for time-varying effects on the outcomes \citep{Valente2019}. Second, heterogeneous treatment effects \citep{Chaise2020}. Due to staggered policy adoption, APE represent weighted averages of effects computed using both treated and untreated outcomes, which may not allow for a comparison with my estimates. Re-weighting the estimator according to \citet{Chaise2020} gives, however, similar results and the bias persists. Therefore, bias may be due to effect heterogeneity across (covariate-specific) groups which are ex-ante unknown. RF methods allow for a data-driven way to estimate these groups and detect ex-post the relevant heterogeneities.

\vspace{0.5cm}
\textit{LASSO.}--Similarly to RF, doubly robust LASSO regression estimates the APE after removing the correlation of covariates with outcomes and prices. Yet, differently from RF, it assumes causal effect homogeneity and, typically, linearity in covariates \citep{athey2017,Belloni2017}. I find that my APE estimates are almost insensitive to these assumptions when differences in covariates are effectively adjusted. Yet, while RF allow to consistently estimate unit level causal effects and identify effect heterogeneity \citep{atheyGRF}, doubly robust LASSO regression is suited for estimation of average treatment effects \citep{Belloni2014}.

\vspace{0.5cm}
\textit{Summary.}--Table  \ref{tab:APEmeth} summarizes all APE estimates by estimation method. Residualized reg via RF is my main estimation method; residualized reg via RF (bin.) assumes no heterogeneity in prices; dynamic fixed effects (FE) assumes constant APE and selection due to time-constant variables; double selection and residualized LASSO reg assume constant APE and linearity after effectively removing differences in covariates.

\renewcommand{\arraystretch}{1.2}

\begin{table}[H]
\centering \footnotesize
\begin{tabular}{llllllllll}
  \hline
  Method/APE estimates & UW  kg & s.e. & RW kg & s.e. & TW kg & s.e. \\
  \hline
residualized reg via RF  & -11.50 & 1.50 & 8.10 & 2.20 & -2.70 & 0.90 \\ 
  residualized reg via RF (bin.) & -16.50 & 1.10 & 13.0 & 2.20 & -3.20 & 2.90 \\ 
 dynamic FE reg  & -7.40 & 0.30 & 2.70 & 0.70 & -4.70 & 0.50 \\ 
   dynamic FE reg (re-weighted)  & -9.06 & 0.98 & 3.96 & 0.76 & -5.10 & 0.40 \\ 
  double selection LASSO reg & -9.30 & 0.40 & 9.40 & 0.60 & -1.00 & 0.50 \\ 
   residualized LASSO reg & -9.00 & 1.30 & 8.80 & 1.00 & -1.00 & 1.30 \\ 
   \hline
\end{tabular}
 \caption{\centering \footnotesize Method comparison: APE estimates in the third policy year. Estimates represent average kg changes in waste for a one cent price increase.}
  \label{tab:APEmeth}  \footnotesize
\end{table}

\end{document}